%% file: pai.tex
\PassOptionsToPackage{hyphens}{url}

\PassOptionsToPackage{table,dvipsnames}{xcolor}

\documentclass[pdflatex,sn-basic]{sn-jnl}

\usepackage{graphicx}
\usepackage{amsmath,amssymb,amsfonts}
\usepackage{textcomp}

\graphicspath{{figures/}{pictures/}{images/}{./}}

\usepackage{enumitem}

\usepackage{svg}

\usepackage[fixed]{fontawesome5}

\usepackage{ellipsis}

\usepackage{tabularx}
\usepackage{multirow}

\usepackage{booktabs}

\usepackage{xspace}

\usepackage{color, colortbl}
\usepackage{xcolor}

\usepackage{subcaption}

\usepackage{soul}

\providecommand{\Description}[2][]{}

\definecolor{principlesBG}{HTML}{c8edc9}
\definecolor{challengesBG}{HTML}{c7e6ef}

\newcommand{\literacychip}[2]{%
  \begingroup
  \sethlcolor{#1}%
  \sffamily\hl{\,#2\,}%
  \endgroup
}

\newcommand{\principleOne}[1]{\literacychip{principlesBG}{#1}}
\newcommand{\principleTwo}[1]{\literacychip{principlesBG}{#1}}
\newcommand{\principleThree}[1]{\literacychip{principlesBG}{#1}}
\newcommand{\principleFour}[1]{\literacychip{principlesBG}{#1}}
\newcommand{\principleFive}[1]{\literacychip{principlesBG}{#1}}

\newcommand{\challengeOne}[1]{\literacychip{challengesBG}{#1}}
\newcommand{\challengeTwo}[1]{\literacychip{challengesBG}{#1}}
\newcommand{\challengeThree}[1]{\literacychip{challengesBG}{#1}}
\newcommand{\challengeFour}[1]{\literacychip{challengesBG}{#1}}
\newcommand{\xloki}{\textsc{xLocEI}\xspace}
\newcommand{\winsert}{\textsc{Wikinsert}\xspace}

\newcommand{\mutualLearning}{\principleOne{\faGraduationCap~Mutual Learning}\xspace}

\newcommand{\futureAlternatives}{\principleTwo{\faRoute~Future Alternatives}\xspace}

\newcommand{\artifactEcologies}{\principleThree{\faLaptop~Artifact Ecologies}\xspace}

\newcommand{\empowermentMediation}{\principleFour{\faFistRaised~Empowerment \& Mediation}\xspace}

\newcommand{\emancipatoryDemocracy}{\principleFive{\faHandshake~Emancipatory Practices \& Democracy}\xspace}

\newcommand{\specialization}{\challengeOne{\faLanguage~Designing for Specialization}\xspace}

\newcommand{\multimodality}{\challengeTwo{\faHandSparkles~Designing for Multimodality \& Distributed Ecosystems}\xspace}

\newcommand{\emergentBehavior}{\challengeThree{\faCodeBranch~Designing for Emergent System Behavior}\xspace}

\newcommand{\humanAugmentation}{\challengeFour{\faBiking~Designing for Human Augmentation}\xspace}

\usepackage{adjustbox}

\newcolumntype{R}[2]{%
    >{\adjustbox{angle=#1,lap=\width-(#2)}\bgroup}%
    l%
    <{\egroup}%
}

\begin{document}


\title[Participatory AI]{Participatory AI: A Democratic Approach to Human-Centered AI}


\author*[1]{\fnm{Niklas} \sur{Elmqvist}}\email{elm@cs.au.dk}

\author[1]{\fnm{Eve} \sur{Hoggan}}

\author[1]{\fnm{Hans-J\"{o}rg} \sur{Schulz}}

\author[1]{\fnm{Marianne} \sur{Graves Petersen}}

\author[1]{\fnm{Peter} \sur{Dalsgaard}}

\author[1]{\fnm{Ira} \sur{Assent}}

\author[1]{\fnm{Olav W.} \sur{Bertelsen}}

\author[1]{\fnm{Akhil} \sur{Arora}}

\author[1]{\fnm{Kaj} \sur{Grønbæk}}

\author[1]{\fnm{Susanne} \sur{Bødker}}

\author[1]{\fnm{Clemens Nylandsted} \sur{Klokmose}}

\author[1]{\fnm{Rachel Charlotte} \sur{Smith}}

\author[1]{\fnm{Sebastian} \sur{Hubenschmid}}

\author[2,1]{\fnm{Christoph A.} \sur{Johns}}

\author[1]{\fnm{Gabriela} \sur{Molina Le\'{o}n}}

\author[1]{\fnm{Anton} \sur{Wolter}}

\author[1]{\fnm{Johannes} \sur{Ellemose}}

\author[1,3]{\fnm{Vaishali} \sur{Dhanoa}}

\author[1]{\fnm{Simon Aagaard} \sur{Enni}}

\author[1]{\fnm{Mille Skovhus} \sur{Lunding}}

\author[1]{\fnm{Karl-Emil Kjær} \sur{Bilstrup}}

\author[1]{\fnm{Juan} \sur{S\'{a}nchez Esquivel}}

\author[1]{\fnm{Luke} \sur{Connelly}}

\author[1,4]{\fnm{Rafael} \sur{Pablos Sarabia}}

\author[4]{\fnm{Morten} \sur{Birk}}

\author[4]{\fnm{Joachim} \sur{Nyborg}}

\author[1]{\fnm{Stefanie} \sur{Zollmann}}

\author[1]{\fnm{Tobias} \sur{Langlotz}}

\author[1]{\fnm{Meredith Siang-Yun} \sur{Chou}}

\author[1]{\fnm{Jens Emil Sloth} \sur{Grønbæk}}

\author[1]{\fnm{Michael} \sur{Wessely}}

\author[1]{\fnm{Yijing} \sur{Jiang}}

\author[1]{\fnm{Caroline} \sur{Berger}}

\author[1]{\fnm{Duosi} \sur{Dai}}

\author[1]{\fnm{Michael Mose} \sur{Biskjaer}}

\author[1]{\fnm{Germ\'{a}n} \sur{Leiva}}

\author[1]{\fnm{Jonas} \sur{Frich}}

\author[1]{\fnm{Eva} \sur{Eriksson}}

\author[1]{\fnm{Kim} \sur{Halskov}}

\author[1]{\fnm{Thorbj{\o}rn} \sur{Mikkelsen}}

\author[1]{\fnm{Nearchos} \sur{Potamitis}}

\author[1]{\fnm{Michel} \sur{Yildirim}}

\author[1]{\fnm{Arvind} \sur{Srinivasan}}

\author[5]{\fnm{Jeanette} \sur{Falk}}

\author[6]{\fnm{Nanna} \sur{Inie}}

\author[1]{\fnm{Ole Sejer} \sur{Iversen}}

\author[1]{\fnm{Hugo} \sur{Andersson}}

\author[1]{\fnm{Midas} \sur{Nouwens}}

\affil*[1]{\orgname{Aarhus University}, \orgaddress{\city{Aarhus}, \country{Denmark}}}

\affil[2]{\orgname{University of Oldenburg}, \orgaddress{\city{Oldenburg}, \country{Germany}}}

\affil[3]{\orgname{TU Wien}, \orgaddress{\city{Vienna}, \country{Austria}}}

\affil[4]{\orgname{Cordulus}, \orgaddress{\city{Aarhus}, \country{Denmark}}}

\affil[5]{\orgname{Aalborg University}, \orgaddress{\city{Aalborg}, \country{Denmark}}}

\affil[6]{\orgname{IT University of Copenhagen}, \orgaddress{\city{Copenhagen}, \country{Denmark}}}


\abstract{\input{content/00-abstract.tex}}



\maketitle

\begin{figure}[b]
    \centering
    \includegraphics[width=\linewidth]{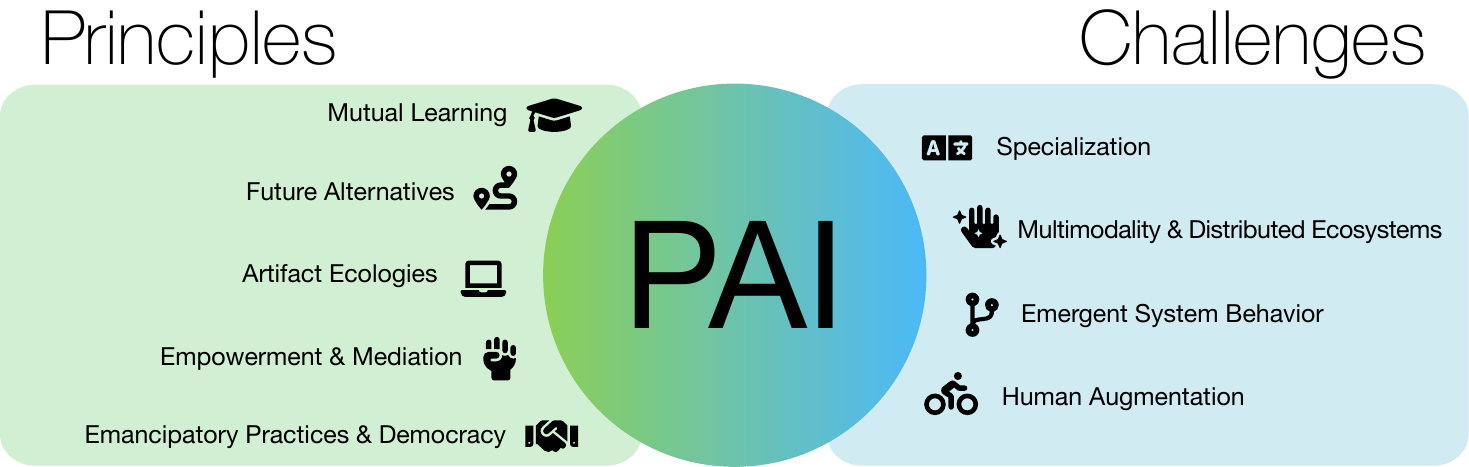}
    \caption{\textbf{Participatory AI.}
    Our framework for Participatory AI (PAI) applies principles from Scandinavian Participatory Design to address challenges that AI systems pose for human agency and democratic participation.}
    \label{fig:teaser}
    \Description{The framework for Participatory AI (PAI), consisting of the five principles; (1) Mutual Learning, (2) Future Alternatives, (3) Artifact Ecologies, (4) Empowerment and Mediation, and (5) Emancipatory Practices and Democracy. The framework also consists of five design challenges of designing for (1) Specialization, (2) Multimodality and Distributed Ecologies, (3) Emergent System Behavior, and (4) Human Augmentation.}
\end{figure}

\input{content/01-intro}
\input{content/02-background}
\input{content/03-pai-framework}
\input{content/04-case-studies}
\input{content/05-discussion}
\input{content/06-research-vision}

\backmatter

\bmhead{Acknowledgements}

\input{content/10-acks}


\bibliography{pai}

\end{document}

%% file: content/00-abstract.tex
AI's transformative impact on work, education, and everyday life makes it as much a political artifact as a technological one. 
Current AI models are opaque, centralized, and overly generic.
The algorithmic automation they provide threatens human agency and democratic values in both workplaces and daily life.
To confront such challenges, we turn to Participatory Design (PD), which was devised in the 1970s to face a similar threat from mechanical automation.
In the PD tradition, technology is seen not just as an artifact, but as a locus of democracy.
Drawing from this tradition, we propose \textit{Participatory AI} as a PD approach to human-centered AI that applies five PD principles to four design challenges for algorithmic automation.
We use concrete case studies to illustrate how to treat AI models less as proprietary products and more as shared sociotechnical systems that enhance rather than diminish human agency, human dignity, and human values.

%% file: content/01-intro.tex
\section{Introduction}

In 1970s Scandinavia, a distinctive approach to technological change emerged amid strong labor unions, robust worker protections, and progressive politics~\citep{Nygaard1978}.
Rather than blindly resisting the technologies that were infiltrating their workplace, workers established democratic processes to influence how mechanical automation was implemented in practice~\citep{Ehn1988}.
This \textit{Participatory Design} (PD) movement recognized workers' democratic right to shape the tools transforming their labor~\citep{EhnSandberg1983, DBLP:series/synthesis/2022Bodker, Nygaard1978}.
The tradition embodied respect for the dignity of the worker and insisted that technology must serve humans, never the reverse~\citep{Ehn1988, DBLP:series/synthesis/2022Bodker}.
In this way, it promoted the expertise of the workers themselves~\citep{DreyfusAndDreyfus1988, Suchman1987}.
Because workers had influence over the implementation of technology, there was less resistance to change~\citep{Nygaard1978}.
PD has since dispersed into many forms; in this paper we return to its roots.

The conditions that gave rise to participatory design have now returned in new form.
Today we face similar challenges as artificial intelligence (AI) automates intellectual rather than physical labor throughout society~\citep{DBLP:journals/ijhci/Shneiderman20, Shneiderman2022}.
While the nature of automation has shifted, familiar tensions persist:
Who owns the work that is to be automated?
What are the values embedded in the automation?
And how is human agency preserved in the face of automation~\citep{DBLP:series/synthesis/2022Bodker, Shneiderman2022}?
As AI systems reshape workplaces, civic institutions, and personal environments, these democratic principles of participatory design must again rise~\citep{DBLP:journals/tochi/BodkerK18}.
And, as before with participatory design~\citep{DBLP:conf/nordichi/Bodker06}, the changes are not restricted to the workplace, but transcend into everyday life.
These democratic ideals, born from workplace struggles yet applicable beyond them, offer guidance for ensuring that AI enhances rather than diminishes human agency and dignity in society as a whole~\citep{DBLP:series/synthesis/2022Bodker, Shneiderman2022, Cuizon2024}.

In this paper, we propose \textit{participatory AI} (PAI) as a Participatory Design (PD) approach to human-centered AI (HCAI)~\citep{Shneiderman2022} that extends democratic design principles to the challenges of algorithmic automation.
While the term has already been discussed in early work involving creative professionals using generative AI~\citep{DBLP:conf/chi/InieFT23}, as well as in critical examinations of power dynamics and participation across AI development~\citep{DBLP:conf/eaamo/BirhaneIPDEGM22}, our treatment interprets the automation of cognitive work through the collective-resource tradition of PD and derives a framework from that.
Our participatory AI framework is built on a synthesis of core PD principles with emerging AI design challenges (Figure~\ref{fig:teaser}), providing actionable guidance for practitioners seeking to implement more democratic approaches.
To illustrate the framework, we present five case studies demonstrating how participatory methods can reshape AI development across domains such as schools, creativity, online knowledge, agriculture, and manufacturing.
Finally, we outline a research agenda that addresses critical questions at this intersection, suggesting directions for scholarship that can strengthen the theoretical foundations of participatory approaches to AI while responding to urgent practical needs.

%% file: content/02-background.tex
\section{Background}
\label{sec:background}

To understand how participatory design principles can address contemporary AI challenges, we examine two parallel technological trajectories that have shaped human-computer interaction over the past seven decades.
The first concerns the fundamental tension between artificial intelligence and intelligence augmentation: two competing visions for how computing should relate to human cognition and agency.
The second traces the emergence of participatory design in 1970s Scandinavia, where democratic design principles were developed specifically to address mechanical automation's threat to human work and dignity.
We also survey recent attempts to bring these two trajectories together under the banner of ``participatory AI.''

\subsection{Scope}

The term ``AI'' covers a wide spectrum of systems, from reinforcement learning agents in scientific research to internet-scale foundation models trained at costs exceeding a billion dollars.
Situated participation in the training of such foundation models faces barriers of capital and scale that no community-based process can realistically overcome.
Our focus is therefore largely on AI systems built \textit{on top of} foundation models; fine-tuned, specialized, or otherwise adapted for specific domains and communities of practice.
This is the layer where participatory design has the most traction: where domain knowledge shapes training data, where interaction norms are negotiated, and where the fit between AI capabilities and local practices is determined.

Our case studies all operate at this level.
The framework we propose, however, also surfaces questions about the foundation layer---questions of data provenance, value embedding, and democratic governance---that we raise without claiming to resolve.

\subsection{Artificial Intelligence vs.\ Intelligence Augmentation}

The field of computing has pursued two fundamentally different visions since its inception.
The so-called Dartmouth Workshop in 1956 established the ambitious goal of making \textit{``every aspect of learning or any other feature of intelligence [...] so precisely described that a machine can be made to simulate it''}~\citep{DBLP:journals/aim/McCarthyMRS06, NewellSimon1976, Crevier1993}.
This autonomous AI vision treats human intelligence as a target for replication and eventual replacement by machines.

Simultaneously, cybernetics explored a different path.
Ashby's feedback systems~\citep{Ashby1956IntroductionToCybernetics} conceptualized intelligence as emerging from human-machine interaction rather than machine autonomy.
Licklider formalized this as \textit{man-computer symbiosis}, arguing that ``\textit{the hope is that, in not too many years, human brains and computing machines will be coupled together very tightly, and that the resulting partnership will think as no human brain has ever thought}''~\citep{Licklider1960ManComputerS}.
Engelbart extended this vision into systematic \textit{intelligence augmentation} (IA), developing frameworks for ``augmenting human intellect'' through technological tools that enhance rather than replace human cognition~\citep{Engelbart1962}.

Until recently, the two camps have evolved in parallel but separately.
Jordan argues that current AI systems pursue ``human-imitative AI'' that mimics human behavior without understanding human needs or social contexts~\citep{Jordan2018}.
Foundation models exemplify this trajectory: massive systems trained on internet-scale data with minimal human oversight~\citep{Bommasani2021}.
These models centralize enormous computational power in corporate hands, making fundamental decisions about information access, knowledge representation, and reasoning processes without democratic input.

The response has been largely technical rather than political.
Li advocates for AI systems that \textit{``augment rather than replace human capabilities''}~\citep{Li2018}, but these approaches treat humans as end-users rather than co-designers of AI systems.
Explainable AI (XAI) attempts to make black-box models interpretable~\citep{DBLP:series/lncs/Samek2019, DBLP:journals/air/MinhWLN22, rudin2019stop}, while Human-Centered AI (HCAI) frameworks emphasize usability and reliability~\citep{Shneiderman2022, DBLP:journals/ijhci/Shneiderman20}.
These technical responses, while valuable, do not address the underlying political question of who should control AI development and for whose benefit.

\subsection{A Brief History of Participatory Design}
\label{sec:history}

According to \citet{DBLP:series/synthesis/2022Bodker}, participatory design can be summarized as including \textit{``activities where users, designers and researchers collaborate towards shared goals."}
\textit{Mutual learning} is one of the core PD principles, emphasizing that designers have to learn about users' current work practices while users learn about technological opportunities~\citep{InspirationCards}.
Mutual learning between these groups is hence important, as is the emphasis that Participatory Design starts with the \textit{current practices} of people in groups and organizations and uses \textit{future alternatives} for joint reflection and action in critical and inclusive ways.

Mutual learning is not just a simple relation between individuals.
PD focuses on \textit{empowerment of people} not only as individuals but as part of their groups and communities, both as they are currently established and for the (possible) future.
Whether in the workplace or elsewhere, such settings are full of conflicts that design as such cannot solve, even though they need to be handled in the processes. 
Seeing human beings as \textit{skillful and resourceful} in the development of their future joint practices sets focus on PD as a set of \textit{emancipatory practices}, also in situations where power is not equally balanced.
In this way, PD is committed to \textit{democracy}.

PD grew out of a number of collaboration projects between researchers and Scandinavian trade unions in the late 1970s~\citep{bodker_cooperative_2000, EhnKyng:1984, BodkerEhnKammersgaard1987, KyngMathiassen1982, Nygaard1978, EhnSandberg1983}.
These projects were initiated to meet growing concerns that computing technologies would replace and deskill human work due to the increased automation of work activities.
Computer-based automation of human work, in particular routine factory and office work, was not a new phenomenon in the late 1970s, but the automation of work that required trained skills or craftsmanship was increasingly being targeted.
In addition, automation moved into the realm of health and care, questioning the quality of work, e.g., in nursing~\citep{BjerknesBratteteig1987}.
PD researchers suggested the tool perspective as an alternative to the dominant system perspective in information systems and HCI research. 
While the system perspective treats technology as an autonomous structure that shapes and constrains human action, the tool perspective emphasizes technology as a human-controlled resource for skilled workers.

The development of critical alternatives became a core value for PD.
As formulated by \citet{Bannon1992}, workers should not be reduced to factors in production machinery, but should be understood as (human) actors that bring knowledge, skills, and agency to their work.
Thus, a critical focus was on how to develop solutions with new technologies that would empower resourceful, active end-users, enhancing democracy and co-determination in the workplace and beyond.
In short, the process of participatory design became an act of empowerment through mutual learning.

PD represents a turn to practice because it treats communities rather than individuals as the fundamental unit for understanding how people use IT-based artifacts.
Hence, interactive technologies should be understood as \textit{instruments} mediating human activity, and motivated in that activity.
While early PD often emphasized the development of a single artifact, most computer-based instruments are now situated in ensembles or \textit{artifact ecologies}~\citep{Jung2008Ecologies, bodker2011ham}. 
In this sense, PD has become a matter of infrastructuring, addressing the local in the context of multiple technologies surrounding local activities, but often embedded within complex hyper-connected global systems~\citep{Karasti2024}.

PD at the methodological level overlaps with other approaches such as co-design, experience-centered design, and interaction design in general.
What makes activities, tools, and techniques distinct in PD, according to \citet{DBLP:series/synthesis/2022Bodker}, is how they are utilized in alignment with the core elements:
To support principles and values of mutual learning, democracy, empowerment of end-users, and to acknowledge the skillfulness of all people involved in the design process.
Working with alternatives is not only an overall goal of PD, but also an ambition that permeates the toolbox and practices in multiple ways~\citep{Simonsen2025}.
At the same time, it is important to emphasize that this level of methods and techniques for direct collaboration between participants is only part of the activities that happen in PD processes.
The hands-on design activities need to be connected with the wider political, democratic, and organizational processes that reach out from the local community in various ways, to policy, national, and global levels, and more, as discussed by \citet{DBLP:journals/tochi/BodkerK18}, and \citet{ComputationalEmpowerment}.

PD was developed in parallel with the early 1980s critique of artificial intelligence~\citep{WinogradFlores1986, Suchman:1988, DreyfusAndDreyfus1988} that pointed to the uniqueness of human practice and expertise, and argued against the possibility of capturing human expertise in so-called expert systems, thereby pointing to actual users as central actors to be supported in technology design.
The need for continued user control becomes emphatic but also more complex as technological ensembles become more opaque and interconnected. 

Over the past decade, PD research, theory, and practice, along with its core values of participation, empowerment, and democracy, have diversified and evolved in novel directions.
Technologies are increasingly dominated by large tech corporations, leading to greater standardization, and challenging engagement.
Participation is challenged by these systems, while also the nature of communities and collectives have become diversified, dispersed, and versatile.
Today, participation unfolds in multiple forms over time, involving different actors, relationships, and interconnected processes across time, geographies, and scales~\citep{Smith2025a, Smith2025b}.
These contemporary conditions are part of the reason for PD's lack of engagement with complex and agentic AI systems, but also the reason why PD needs to be engaged in developing critical AI alternatives~\citep{frauenberger2024emerging}. 

The basic challenges we are facing with today's generative AI are very similar to the challenges faced by early PD where physical work was being automated.
In a world where computer technology is putting human lives and work under pressure~\citep{DBLP:journals/tochi/BodkerK18}, critical questions concerning the role of human beings, agency, skills, and control are central.
Today's accelerated AI development requires new partnerships with users and stakeholders to democratically govern how these technologies integrate into existing sociotechnical systems.

\subsection{Participation Meets AI}
\label{sec:participation-meets-ai}

The term ``participatory AI'' has surfaced in recent years as researchers recognize the need for democratic involvement in AI development~\citep{DBLP:conf/chi/ZytkoWGBL22}.
This emerging field encompasses diverse approaches, from legal requirements for civil society consultation to methods for inclusive data labeling and co-design~\citep{DBLP:journals/firstmonday/YoungESTGHM24}.
Yet while these works share a commitment to participation, they struggle with implementation, particularly when confronting the scale and complexity of contemporary AI.

A fundamental tension exists between the localized engagement characteristic of participatory methods and the globalized operation of commercial AI.
\citet{DBLP:journals/firstmonday/YoungESTGHM24} identify three fault lines where participatory and commercial AI logics diverge: centralized versus distributed development, calculable versus self-identified publics, and instrumental versus intrinsic perceptions of the value of public input.
\citet{SureshTsengYoung2024} identify a ``participatory ceiling'' limiting meaningful community control over foundation models, proposing organizational workarounds through domain-specific intermediate layers but stopping short of a comprehensive design process.

\citet{DelgadoYangMadaio2023} provide the most comprehensive analysis of this ``participatory turn,'' synthesizing literature across technology design, political theory, and the social sciences into a conceptual framework for evaluating participatory approaches.
Their empirical investigation, combining analysis of published research with interviews of AI practitioners, reveals that most participatory efforts occur late in development cycles after core architectural decisions have already been made.
They also find enormous variation in how participation is conceptualized, with implicit disagreements about whether participants should have substantive agency or merely provide input to designer-controlled processes.

Several research efforts have attempted to bridge these gaps.
\citet{DBLP:conf/eaamo/BirhaneIPDEGM22} identify opportunities for community participation in AI development, arguing that those most affected by AI systems should be centered in development processes.
Their framework, however, lacks concrete methods for direct involvement.
\citet{SloaneMossAwomolo2022} provide an important critique of superficial participation, warning against ``participation washing'' where community involvement becomes extractive rather than empowering.
They distinguish between participation as work, consultation, and justice, arguing that current practices often exploit rather than empower communities.
\citet{DBLP:conf/eaamo/CorbettDE23} sharpen this critique by applying Arnstein's Ladder of Citizen Participation to recent participatory AI scholarship, finding that most work informs or consults rather than partners with or delegates control to participants: the lower rungs of meaningful participation.
Inie et al.\ apply participatory methods specifically to generative AI tools for creative professionals, examining concerns and expectations about AI systems that might automate or augment creative work~\citep{DBLP:conf/chi/InieFT23}.
While their approach represents genuine participatory engagement, it is limited to a single domain.

More promisingly, some projects demonstrate that participatory research can scale.
\citet{DBLP:conf/emnlp/NekotoMMFFAMKOS20} developed machine translation systems for over 30 African languages through participatory research methods, enabling contributors without formal training to make scientific contributions while producing novel datasets and benchmarks.
Their work shows that participation need not be confined to small-scale, localized interventions.
Other work explores domain-specific co-design: \citet{DBLP:conf/chi/YildirimZSKBARD24} report on multidisciplinary workshops for AI concept ideation in healthcare, while \citet{DBLP:conf/fat/TsengYQRS25} co-design a journalist-controlled LLM, surfacing tensions between ``one-size-fits-all'' foundation models and professional practices.
Both identify the gap between what technologists build and what domain experts need.

Yet the field remains fragmented.
\citet{DBLP:journals/firstmonday/YoungESTGHM24} argue that scaling participation requires infrastructural investment comparable to what has enabled commercial AI to scale.
Just as scaling commercial AI demanded significant resources, scaling participation will require dedicated infrastructure for the practical dimension of shifting power.
This observation points toward the contribution of this paper: rather than treating participation as an add-on to existing AI development, we propose a framework that integrates participatory design principles throughout the AI lifecycle, drawing on PD's experience with democratic technology development.

%% file: content/03-pai-framework.tex
\section{Participatory AI -- Framework}
\label{sec:framework}

The introduction of large AI models has fundamentally changed software both in scale and capability but also with regards to epistemology, agency, and embeddedness in feedback loops and power structures.
Therefore participatory design must evolve to adapt to these changing conditions.
\textit{Participatory AI} (PAI) aims to make AI answerable to the communities it shapes through the application of principles from Participatory Design (Figure~\ref{fig:teaser}).
It insists that those who use the potentials of AI or live with its consequences should have a say in its conception, training, deployment, and evolution.
In PAI, AI systems are considered dynamic, learning systems embedded in practice whose design cannot be finalized up front but must be negotiated, monitored, and adapted in collaboration with those it affects.
Here we first describe the method we followed to develop the framework, and then present its principles and design challenges.

\subsection{Method}
\label{sec:framework-method}

Our participatory AI framework emerged through an iterative process combining theoretical analysis with empirical engagement over approximately two years.
We began from the premise that participatory design must evolve to address AI's distinctive characteristics, but that the specific adaptations required would become visible only through sustained engagement with AI in practice.

The process began with a special interest group convened within our university during fall 2024, bringing together researchers with expertise spanning human-computer interaction, AI systems, and participatory design.
This group identified initial tensions between established PD principles and emerging AI development practices.
During spring, summer, and early fall 2025, we held seven workshops to collect examples, case studies, relevant literature, and candidate design challenges from participants' ongoing research.
Smaller workgroups formed at different times during the year to develop specific components of the framework independently, some focusing on translating PD principles to AI contexts, others on articulating design challenges that surfaced across domains.
Synthesis occurred during fall 2025, where we consolidated the workgroups' outputs.

The five PAI principles (Section~\ref{sec:pai-principles}) were derived directly from established participatory design literature~\citep{DBLP:series/synthesis/2022Bodker, SimonsenRobertson2013, DBLP:journals/tochi/BodkerK18}.
We selected principles that have remained central to PD across its evolution: mutual learning as the foundation of designer-user collaboration, future alternatives as PD's commitment to exploring multiple trajectories, artifact ecologies as the lens for understanding technology-in-context, empowerment and mediation as the dual focus on human agency and tool perspective, and emancipatory democracy as PD's political commitment.
We claim no novelty for the principles themselves; our contribution is their reinterpretation for AI.

The four design challenges (Section~\ref{sec:design-challenges}) emerged differently.
Drawing on the authors' collective research across education, creative industries, online knowledge systems, agriculture, and manufacturing, we identified recurring sites of tension where AI systems posed difficulties not adequately addressed by existing PD frameworks.
Through iterative discussion, we consolidated these into four challenges that capture where AI diverges from traditional software in ways that matter for participatory practice:
\textit{specialization} addresses how AI behavior is shaped implicitly through data rather than explicit specification;
\textit{multimodality} addresses how distributed data flows create opacity and raise questions of consent and ownership;
\textit{emergent behavior} addresses how probabilistic systems produce unpredictable outputs requiring ongoing negotiation of acceptability;
and \textit{human augmentation} addresses the tension between automation and human agency in AI-mediated work.

The proposed framework is in its third iteration.
An initial straw man (May 2025) enumerated nine parallels between the automation PD confronted in the 1970s and AI today; the framework workgroup consolidated these into seven candidate principles anchored in the four PD commitments of \citet{DBLP:series/synthesis/2022Bodker}, and consensus discussion reduced them to five.
Three candidates did not survive as separate principles: \textit{agency and control} and \textit{socio-technical grounding} were merged into empowerment and mediation, since the tool perspective already carries both; \textit{infrastructuring} was folded into artifact ecologies; and \textit{measuring success} was removed because it names an outcome of participation rather than a commitment.
\textit{Transparency}, present in the straw man, moved to the AI side of the framework, where it reappears as opacity under the multimodality challenge.

The design challenges went through a parallel reduction.
We first planned to cross the PD commitments with a five-stage AI life cycle (inception, training, use, updating, retirement) but dropped the life-cycle axis in August 2025 when it added little beyond what the challenges already captured; a candidate challenge on \textit{centralization and heterogeneous computing environments} was likewise set aside, and \textit{agentic AI} was absorbed into human augmentation.
We retained a challenge when it recurred across the domains represented in the group, bore directly on participation rather than on technical performance, and named a way in which AI departs from conventional software.

We do not claim these four challenges are exhaustive.
Other framings are possible, and additional challenges---such as sustainability, labor equity in data annotation, or regulatory compliance---merit attention~\citep{Crawford2018}.
Our selection reflects challenges that surfaced consistently across diverse domains and that foreground questions of participation, agency, and democratic governance.
The case studies in Section~\ref{sec:cases} illustrate how these principles and challenges manifest in practice.
We use them to demonstrate the framework's analytical utility; they were selected from the group's broader research after the challenges had been consolidated, not used to derive them.

\subsection{Positionality Statement}

Our author team represents a large and diverse group of academic researchers spanning Europe, North America, South America, and Asia, with ages ranging from 20s to 60s.
The team includes both early-career researchers (Ph.D. students and postdocs) and faculty members in roughly equal proportions, along with a small contingent of industry representatives from an AI startup.
Gender distribution is approximately two-thirds men and one-third women.

We approach this work as adherents to the Scandinavian participatory design tradition, though to varying degrees based on our individual research trajectories and geographic contexts.
Many of our team members are recognized authorities within this tradition, bringing decades of experience in participatory methods and democratic design principles.
Even those authors who originate from other parts of the world operate within the Nordic cultural and academic context.

Our disciplinary backgrounds span human-centered computing and computer science, including HCI, data visualization, education technology, large language models and AI, data-intensive systems, databases, interactive systems, interaction design for children, manufacturing systems, extended reality (XR/AR/VR), human-centered AI, communication studies, participatory design methodology, and labor union studies. 
This breadth reflects our commitment to examining participatory AI across diverse application domains.

Politically, our team tends toward progressive, liberal, and democratic values, though we acknowledge the difficulty of characterizing such a large group definitively.
We recognize that our shared commitment to participatory design principles reflects particular political and ethical positions about technology's role in society: namely, that democratic participation should guide technological development rather than being an afterthought.
We also acknowledge the special political and social circumstances of our host country: deeply embedded democratic traditions, strong trust in government institutions, robust labor union representation, comprehensive social safety nets, and a cultural emphasis on collective decision-making and consensus-building.

We are aware that our backgrounds, experiences, and beliefs introduce biases that favor participatory approaches and may lead us to underemphasize competing perspectives on AI development.
Our deep investment in the Scandinavian PD tradition, while providing valuable expertise, may also limit our ability to recognize alternative frameworks for democratic technology design that emerge from different cultural contexts.
We address these limitations in Section~\ref{sec:discussion}.

\subsection{PAI Principles}
\label{sec:pai-principles}

Here we cast \textit{participatory design principles} onto AI:

\begin{description}

    \item[I.] \mutualLearning{} --
    Mutual learning in PD is more than simple knowledge transfer; it is about establishing spaces where diverse forms of expertise are valued and exchanged.
    In AI contexts, this includes learning between communities of users and designers, and jointly developing new literacies about how AI systems adapt to user preferences and partialities, and how designers and communities of users might take this into account.

    \item[II.] \futureAlternatives{} --
    PD rejects the notion that technological trajectories are predetermined.
    In AI development, this principle manifests as a commitment to exploring multiple possible futures through collaboration and critical reflection.
    Rather than accepting AI as an inevitable force that communities must adapt to, AI design processes must be a space for future alternatives where different visions should be explored and debated.

    \item[III.] \artifactEcologies{} --
    Understanding AI systems as components within broader artifact ecologies acknowledges that these technologies rarely operate in isolation.
    They enter environments already rich with tools, practices, infrastructures, and sociotechnical practices.
    Artifact ecologies~\citep{bodker2011ham, korsgaard2022collectives} help us to take into account how AI will interact with and transform existing sociotechnical configurations.

    \item[IV.] \empowermentMediation{} --
    Empowerment in PD extends beyond individual skill development to encompass the strengthening of communities and their collective ability to shape their technological environments.
    Furthermore, interactive technologies should be understood as instruments, mediating human activity, and motivated in that activity.
    For AI systems, this means designing not just for individual users but for groups, organizations, and communities as units of empowerment.

    \item[V.] \emancipatoryDemocracy{} --
    PD's commitment to democracy recognizes that technology design is inherently political, involving groups and communities with conflicting interests and power, serving certain values and interests, possibly at the cost of others. 
    Emancipatory practices in AI development can support interests and democratize decisions, especially in contexts where power imbalances might otherwise exclude certain voices.

\end{description}

\subsection{Design Challenges}
\label{sec:design-challenges}

To create an AI system, conscious design decisions need to be made in the curation and synthesis of data that appropriately captures the desired system behavior.
This property permeates all aspects of AI systems, and we group its concerns under the following four fundamental \textit{design challenges} of AI, ranging from the inception of AI systems, their design over time, to their continued use.

\begin{description}

    \item[A.] \specialization{} --
    AI systems are rarely built from scratch; they are typically fine-tuned from foundation models whose broad capacities are narrowed toward specific tasks~\citep{Bommasani2021, Ouyang2022}.
    Unlike traditional programming, where desired behaviors are explicitly encoded, AI specialization works indirectly: behavior is shaped through training data, model parameters and architecture, prompt engineering, reinforcement from human feedback~\citep{Mei2025}.
    This indirection poses two distinct challenges for participation.
    First, the implicit nature of behavioral specification makes it difficult for non-experts to understand, predict, or contest system outputs.
    Second, decisions about what constitutes ``desired'' behavior embed values that may not reflect the plurality of perspectives among affected communities.
    Participatory involvement in optimization---through iterative feedback on outputs, co-curation of training data, or deliberation over behavioral targets---can address both challenges: making AI behavior more legible while ensuring that the values embedded reflect broader input.

    \item[B.] \multimodality{} -- 
    AI systems increasingly process and generate multiple data types---text, images, audio, video, sensor readings---across distributed networks of devices and services~\citep{Bordes2024, Bommasani2021, DBLP:journals/csur/SeabornMPO21}.
    This multimodal, distributed architecture creates two challenges for participation.
    First, opacity: users often cannot trace how their data moves through these systems, where it is stored, or how it is transformed~\citep{Crawford2018}.
    As models internalize data into weights, the boundary between input and system becomes blurred, making questions of data ownership, rights, control, and confidentiality difficult to resolve.
    Second, the intimacy of certain modalities raises distinct concerns: biometric inputs like voice, face, or gesture are both personally identifying and vulnerable to misuse.
    Participatory AI addresses these challenges by insisting that communities make informed decisions about which data channels to expose, under what conditions, and for what purposes.
    This requires not only technical transparency but negotiated consent---ongoing dialogue about acceptable data practices rather than one-time permissions.
    \textit{Co-creation} practices~\citep{Sanders2008CoCreation}, where communities actively shape system boundaries, can foster accountability while balancing data richness against privacy and autonomy.

    \item[C.] \emergentBehavior{} -- 
    AI systems built on probabilistic models produce outputs that can vary even with identical inputs and may exhibit unintended behaviors not anticipated by designers~\citep{Guo2025, CampoloSchwerzmann2023, DBLP:conf/iclr/WeiBZGYLDDL22, Marks2025}.
    This unpredictability distinguishes AI from traditional software, where bugs are typically reproducible and traceable.
    Emergence creates a downstream challenge: because system behavior cannot be fully specified in advance, communities must make ongoing decisions about what constitutes acceptable performance.
    What counts as a ``good'' output depends on context and values that may differ across user groups and change over time.
    Participatory practices become essential not only for identifying when systems behave unexpectedly, but for negotiating the standards against which behavior is judged.
    Feedback loops~\citep{Liu2023, Weidinger2025} serve this function, but their design---metrics, interpretation, and decision-making---is itself a site requiring participation.

    \item[D.] \humanAugmentation{} -- 
    Unlike traditional software operated through explicit commands, AI systems are shaped through data: prompts, uploaded media, sensor readings, retrieved documents, and tool outputs collectively influence behavior in real time~\citep{Mei2025}.
    This reframes interaction as a participatory data practice; users co-constitute system behavior through what they provide.
    Yet this also surfaces a fundamental tension between human oversight and system autonomy.
    More automation reduces the burden of interaction but may also diminish opportunities for users to guide, correct, or learn from the system~\citep{DBLP:journals/ijhci/Shneiderman20}.
    Conversely, maintaining human control requires effort that may undermine the efficiency gains AI promises.
    Where to draw this line is not a technical question but a participatory one: different communities, tasks, and contexts will warrant different balances.
    Training data defines a system's foundational capacities~\citep{Penedo2024, Li2024}, while interaction data shapes its ongoing behavior~\citep{DBLP:conf/nips/BrownMRSKDNSSAA20}; both are sites for participatory intervention.
    Participatory AI demands that users help define what constitutes appropriate input, when delegation to AI is acceptable, and how to audit system responses, ensuring that AI augments human expertise rather than displacing it.

\end{description}

Together, these four design challenges reveal that AI systems are not designed once and for all.
They are shaped by data, evolve through use, and embed shifting norms and existing power relations.
Yet they capture only a fraction of the design decisions to be made and the trade-offs to be negotiated.
As described by \citet{Crawford2018}, there are many more issues besides a system's technical capabilities and the users' preferences, such as questions about sustainability (i.e., how much compute power and natural resources to use), legality (i.e., how much leeway to give on using copyrighted and personal training data), equity (i.e., how to ensure fair compensation for manual labor involved), and safety (i.e., how strictly to moderate AI generated contents).
Participatory practices can serve as a framework on these levels of AI design as well, by involving communities and ensuring a democratic and ethical process in finding suitable compromises.

%% file: content/04-case-studies.tex
\section{Case Studies}
\label{sec:cases}

Here we present five case studies that illustrate how participatory design principles apply to contemporary AI systems across diverse domains.
These cases serve as concrete examples of our participatory AI framework in practice, demonstrating both successful applications of democratic design principles and missed opportunities where traditional AI development approaches limited meaningful participation.
Their purpose is analytical rather than empirical.
We explicitly did not derive our participatory AI framework from these cases; instead, we use them to illustrate how the framework can guide analysis and description of AI systems through a participatory lens. 
Each case reveals different aspects of the tension between algorithmic automation and human agency, showing how participatory approaches can reshape AI development to enhance rather than diminish democratic participation.
Table~\ref{tab:case-studies} provides an overview of how each case study addresses different principles and challenges within our framework.
Each case opens with a short summary, followed by the full account, including figures and study details.

\begin{table}[htb]
    \centering
    \caption{\textbf{Case studies.} Overview of case studies and which parts of the PAI framework they address.}
    \label{tab:case-studies}
    \begin{tabular}{@{}c@{}}
    \includegraphics[width=\textwidth]{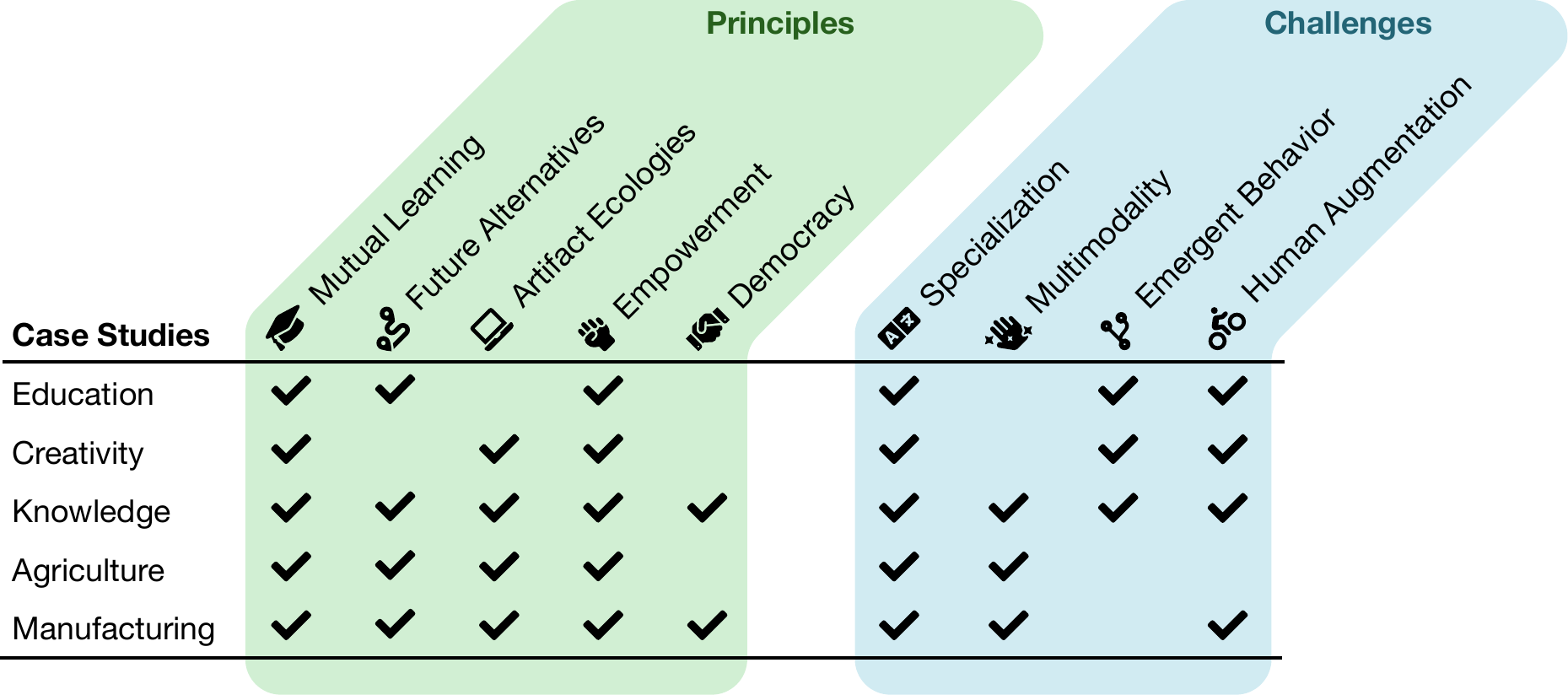}
    \end{tabular}
    \Description{
    An overview of which principles and challenges the five case studies address. The Education case study address the principles of Mutual Learning, Future Alternatives, and Empowerment. The challenges are Specialization, Emergent Behavior, and Human Augmentation. The Creativity case study address the principles of Mutual Learning, Artifact Ecologies, and Empowerment. The challenges are Specialization, Emergent Behavior, and Human Augmentation. The Online Knowledge case study address all the principles and all the challenges. The Agriculture case study address the principles of Mutual Learning, Future Alternatives, Artifact Ecologies, and Empowerment. The challenges are Specialization, and Multimodality. The manufacturing case study address all the principles, with the challenges being Specialization, Multimodality, and Human Augmentation.  
    }
\end{table}

\subsection{Case Selection}

We selected case studies using three inclusion criteria.
First, all cases had to be \textbf{familiar} to at least one of the authors, following a purposeful sampling approach that prioritizes depth of understanding over statistical representation.
Second, each case had to include some form of \textbf{participatory design elements}, whether through collaborative development processes, user involvement in system design, or community-centered implementation strategies. 
Third, all cases involve \textbf{AI or algorithmic automation} that directly impacts human work practices or decision-making.

We acknowledge that this approach produces an illustrative rather than exhaustive sample.
Our selection reflects the research networks and domains accessible to our author team, potentially under-representing certain communities and application areas.
However, this limitation serves our analytical purpose: we seek to demonstrate how participatory design principles can illuminate both opportunities and challenges in AI development across varied contexts.
As shown in Table~\ref{tab:coverage}, our selection spans several sociotechnical dimensions: work types range from creative and knowledge work to physical production; organizational scales include individual practitioners, small teams, volunteer communities, and industrial organizations; and AI roles vary from generative collaboration to decision support and perceptual augmentation.
While not exhaustive---notably absent are healthcare, finance, and government services---the cases collectively cover public, private, and commons-based sectors with distinct governance structures and participatory challenges.

\begin{table}[htbp]
    \centering
    \scriptsize
    \caption{\textbf{Case study coverage.}
        While constrained by our inclusion criteria (familiarity, participatory design elements, and automation), our case studies have good coverage across sociotechnical dimensions.}
    \label{tab:coverage}
    \setlength{\tabcolsep}{4pt}
    \begin{tabular}{lp{3.0cm} p{2.5cm} p{2.5cm} l}
        \toprule
        \textbf{Case Study} & \textbf{Work Type} & \textbf{Organizational Scale} & \textbf{AI Role} & \textbf{Sector} \\
        \midrule
        Education & Cognitive/service & Small teams & Pedagogical tool & Public \\
        Creativity & Creative/cognitive & Individual practitioners & Generative collaborator & Private \\
        Knowledge & Knowledge/curation & Volunteer community & Infrastructure support & Commons \\
        Agriculture & Hybrid cognitive-physical & Small enterprises & Decision support & Private \\
        Manufacturing & Physical/perceptual & Industrial organization & Perceptual augmentation & Private \\
        \bottomrule
    \end{tabular}
\end{table}

Importantly, we include cases that represent both positive examples of participatory principles and instances where traditional AI development approaches created missed opportunities for meaningful participation.
In other words, no single case study is intended to be elevated into an exemplar, but rather merely to illustrate our framework.
Several cases highlight lessons learned from projects where participatory ideals were compromised by technical constraints, institutional pressures, or insufficient attention to power dynamics.
Our aim is to be upfront with such deficiencies.

\input{content/04.1-case-education}

\input{content/04.2-case-creativity}

\input{content/04.3-case-online-knowledge}

\input{content/04.4-case-agriculture}

\input{content/04.5-case-manufacturing}

\subsection{Cross-Case Synthesis}
\label{sec:synthesis}

Table~\ref{tab:synthesis} summarizes what each case study contributes to our understanding of participatory AI, the central tensions encountered, and the factors that enabled or constrained participatory engagement.
Rather than simply cataloging which principles each case addresses, this synthesis highlights the \textit{mechanisms} through which PAI operated and where it fell short.

\begin{table*}[tbh]
    \centering
    \caption{\textbf{Cross-case synthesis showing how participatory AI operated in each domain.}
    The table identifies the primary mechanism through which PAI contributed value, the central tension encountered, enabling factors, and constraints that limited impact.}
    \label{tab:synthesis}
    \scriptsize
    \begin{tabular}{@{}>{\raggedright\arraybackslash}p{0.8cm}>{\raggedright\arraybackslash}p{2.6cm}>{\raggedright\arraybackslash}p{2.6cm}>{\raggedright\arraybackslash}p{2.6cm}>{\raggedright\arraybackslash}p{2.6cm}@{}}
    \toprule
    \textbf{Case} & \textbf{Primary PAI Mechanism} & \textbf{Central Tension} & \textbf{Enabling Factor} & \textbf{Limiting Factor} \\
    \midrule
    1
    & Fine-tuning as participation: students annotate domain texts to shape model behavior
    & AI literacy requires multidisciplinary expertise that conflicts with siloed school subjects
    & Longitudinal partnership with one school; teachers as co-designers
    & Resource constraints in educational system; scaling beyond pilot school \\
    \addlinespace
    2
    & Diagnostic participation: interviews and prototype evaluation surface where AI undermines professional norms
    & Augmentation boundaries are negotiated in use, not predetermined by design
    & Two complementary methods (30 practitioner interviews + prototype evaluation with designer pairs)
    & No interventional design outcome; participation itself is uneven across freelancers, studios, and marginalized groups \\
    \addlinespace
    3
    & Explicit community choice of augmentation over automation through participation
    & Functional vs.\ full algorithmic transparency; balancing usability with understanding
    & Integration into existing artifact ecology (Wikipedia norms, browser-based editing)
    & Small user study sample; scaling governance across 300+ language editions \\
    \addlinespace
    4
    & Visual design negotiation: probabilistic outputs adapted to farmer expectations
    & Model fidelity vs.\ user mental models; data ownership vs.\ privacy concerns
    & Farmer on advisory board; iterative workshop feedback
    & Farmers don't own their data; benefits favor larger farms with data infrastructure \\
    \addlinespace
    5
    & User rejection as design signal: workers chose table display over AR glasses
    & Researcher-driven technical methods vs.\ user-driven interface design
    & Field studies and future workshops; COVID-era remote prototyping tools
    & Worker access constraints; no longitudinal evaluation; limited democratic scope \\
    \bottomrule
    \end{tabular}
\end{table*}

The cases illustrate four patterns that bear on the framework's practical application.
For each pattern, we also discuss the design and sociotechnical recommendations they suggest (Sections~\ref{sec:d-recommendations} and~\ref{sec:st-recommendations}).

\paragraph{Early participatory investment reduces downstream costs.}

In each case, upfront engagement with stakeholders prevented costly misalignments.
The education project's teacher partnerships transformed potential resistance into ownership.
Wikinsert's participatory dialogue explicitly rejected automation, avoiding development of unwanted features.
The agriculture startup's farmer workshops revealed that probabilistic weather maps---though more accurate---conflicted with established visual conventions, enabling a design compromise before market release.
The manufacturing case showed that workers' rejection of AR glasses, surfaced through future workshops, redirected development toward a more practical table-surface display.
The creativity case adds a cautionary complement: without such early investment, creative professionals are left to learn AI system behavior through trial and error, producing a \textit{structural} mutual learning gap where domain expertise never reaches model developers and technical knowledge never reaches practitioners.
These examples suggest that participatory methods function as a form of risk mitigation: the resources invested in early engagement are offset by avoided rework and resistance.
This pattern motivates our recommendation to reframe AI systems as \textit{situated collaborators} co-designed with domain practitioners (DR1), and to treat cooperative interaction as a site of mutual learning (DR2).

\paragraph{Augmentation requires explicit negotiation, not assumption.}

The distinction between augmentation and automation is not self-evident; it must be actively constructed through participatory processes.
Wikipedia editors explicitly chose tools that scaffold judgment rather than replace it, a choice that emerged from dialogue, not technical default.
In manufacturing, the AI system relieves cognitive burden (identifying near-symmetrical objects that workers ``would never learn to distinguish'') without claiming to replace worker skill.
The education case reframes the question entirely: rather than asking whether AI augments or automates, students participate in \textit{defining} what constitutes good and bad model outputs.
The creativity case provides the most granular evidence for this pattern.
Through prototype evaluations, designers drew clear boundaries between phases where AI-generated concepts were productive (fast-paced divergent ideation) and where they were unwelcome (convergent judgment requiring situational knowledge and personal style).
These boundaries were not predetermined by the interface but emerged through use, confirming that \humanAugmentation is not a property of systems but an ongoing achievement of participatory practice.
This motivates our recommendation to reconfigure practice, power, and ownership around AI tools (STR1), and to defend future alternatives by resisting premature closure on what roles AI should play (STR2).

\paragraph{Technical AI decisions often remain outside participatory scope.}

Despite substantive user involvement in problem identification and interface design, core technical choices---model architectures, training procedures, optimization objectives---typically remained researcher-driven.
The manufacturing case states this explicitly: \textit{``The technical AI methods were chosen solely by the researchers.''}
In agriculture, synthetic training data generation and model architecture decisions occurred without farmer input.
The creativity case extends this observation to the industry level: foundation models are adapted to creative domains with minimal involvement from the practitioners whose traditions and standards they affect, and specialization decisions are made without access to the tacit criteria that govern professional quality.
This produces what interviewees described as \textit{generative flattening}: stylistic convergence driven by shared defaults and common datasets.
The education case offers a partial counterexample, where student annotation directly shapes model specialization, suggesting that \specialization may be a more accessible entry point for participatory technical engagement than architecture or training design.
This gap underscores the need for locally controlled models where communities can influence not just interfaces but underlying system behavior (DR3), and for co-design processes that extend participation into the specialization of foundation models (DR1).

\paragraph{Individual participation is insufficient without collective governance structures.}

The cases collectively reveal that participatory gains are fragile when they depend on individual projects or the goodwill of technology providers.
The creativity case makes this explicit: professional associations, unions, and informal creative communities were identified as necessary mediators for sustaining participatory influence over platform-embedded AI.
Without such structures, questions of authorship, value distribution, and professional standards cannot be addressed through interface design alone; they require governance.
The Wikipedia case demonstrates what such governance can look like in practice: Wikimedia's humans-first AI strategy provides a community-endorsed policy framework within which tools like Wikinsert operate.
The agriculture case shows the inverse: farmers share data with the startup but do not own it, and the benefits of AI-driven decision-making accrue disproportionately to larger operations with more data infrastructure.
The education case suggests a middle path through community-building: teachers uploading datasets and sharing practices represent an emergent form of collective participation, even if formal governance structures are not yet in place.
These observations point toward a need for PAI to engage not only with design processes but with the institutional and economic structures that determine who benefits from AI systems over time.
This motivates treating AI as public infrastructure subject to democratic accountability (STR3), and ensuring that data ownership and value distribution are explicitly negotiated rather than defaulting to platform providers (STR1).

%% file: content/04.1-case-education.tex
\subsection{Case Study 1: Education}
\label{sec:education}

Generative AI has entered secondary classrooms through public platforms that students use freely and teachers cannot control~\citep{Kasneci_chatgpt_for_good_education, Harvey2015_AI_harms_teachers}.
This case concerns a project with first-language teachers in Danish high schools~\citep{Bilstrup2025_AI_integration} whose aim was to let teachers set the terms of AI integration: subject knowledge as the way into AI, and AI methods as a way of working with the subject.

The project team combined high school teachers, educational researchers, NLP researchers, and digital humanities researchers, working through workshops, prototyping, and co-teaching over several months (\mutualLearning).
Its tools expose how language models represent and generate text, and unplugged activities~\citep{Conelly2025_NLP_tools} let classes rehearse the data practices behind AI without depending on systems they do not control.
A longitudinal partnership with one school, followed by workshops for teachers at other schools, seeded a community of teachers who upload datasets and share activities (\empowermentMediation, \futureAlternatives).

We read the project's central mechanism as \specialization{} enacted in the classroom.
Students annotate sentiment in literature they already study, such as H.C.\ Andersen's fairy tales; the teacher leads the class through pre-annotating, annotating, testing, and iterating a model; and datasets aggregated across classrooms train models returned to the contributing classes.
Fine-tuning becomes participation, with subject competence shaping model behavior directly.
Annotating argument quality did the same for \emergentBehavior, opening discussion of how norms enter LLMs through manual labor, and the arrangement as a whole substitutes for zero-shot prompting under \humanAugmentation: students decide what a good example is instead of receiving one.

AI drove subject development here rather than displacing it, but only because teachers came to understand the data practices behind the tools and researchers came to understand the subject.
That shared understanding costs time a strained school system rarely has, and the teacher community has not yet grown beyond the pilot school.

AI tools are rapidly entering classrooms through publicly accessible platforms, creating new realities where students use AI while teachers struggle to maintain agency over their subjects.
This case study examines how first language teachers can gain agency in AI integration, using subject knowledge to engage with AI technologies while empowering both teachers and students to interact with AI on their own terms rather than through generic interfaces.

\begin{figure*}[htb]
    \centering
    \begin{subfigure}[b]{0.32\textwidth}
        \includegraphics[width=\linewidth]{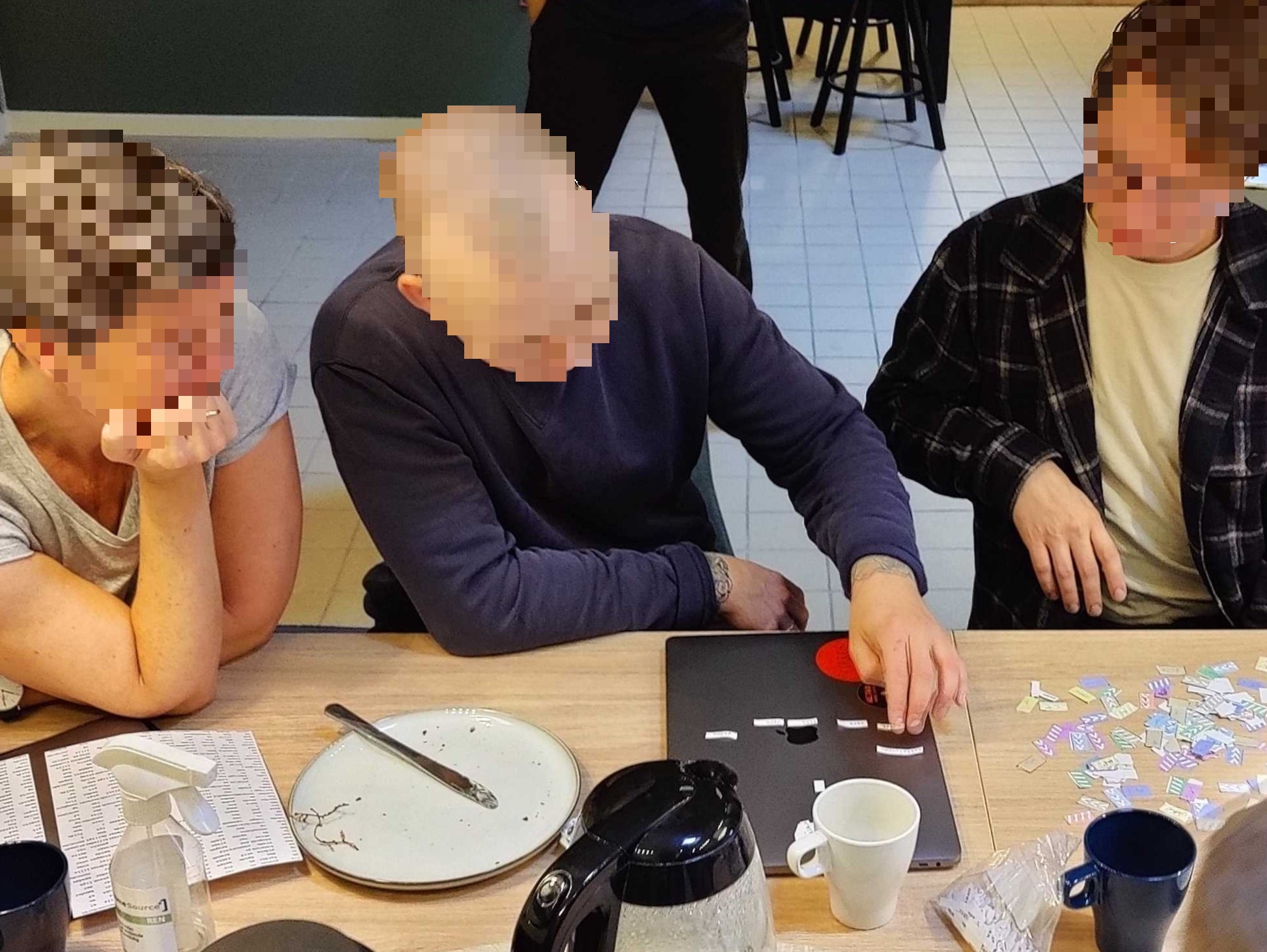}
    \end{subfigure}
    \begin{subfigure}[b]{0.32\textwidth}
        \includegraphics[width=\linewidth]{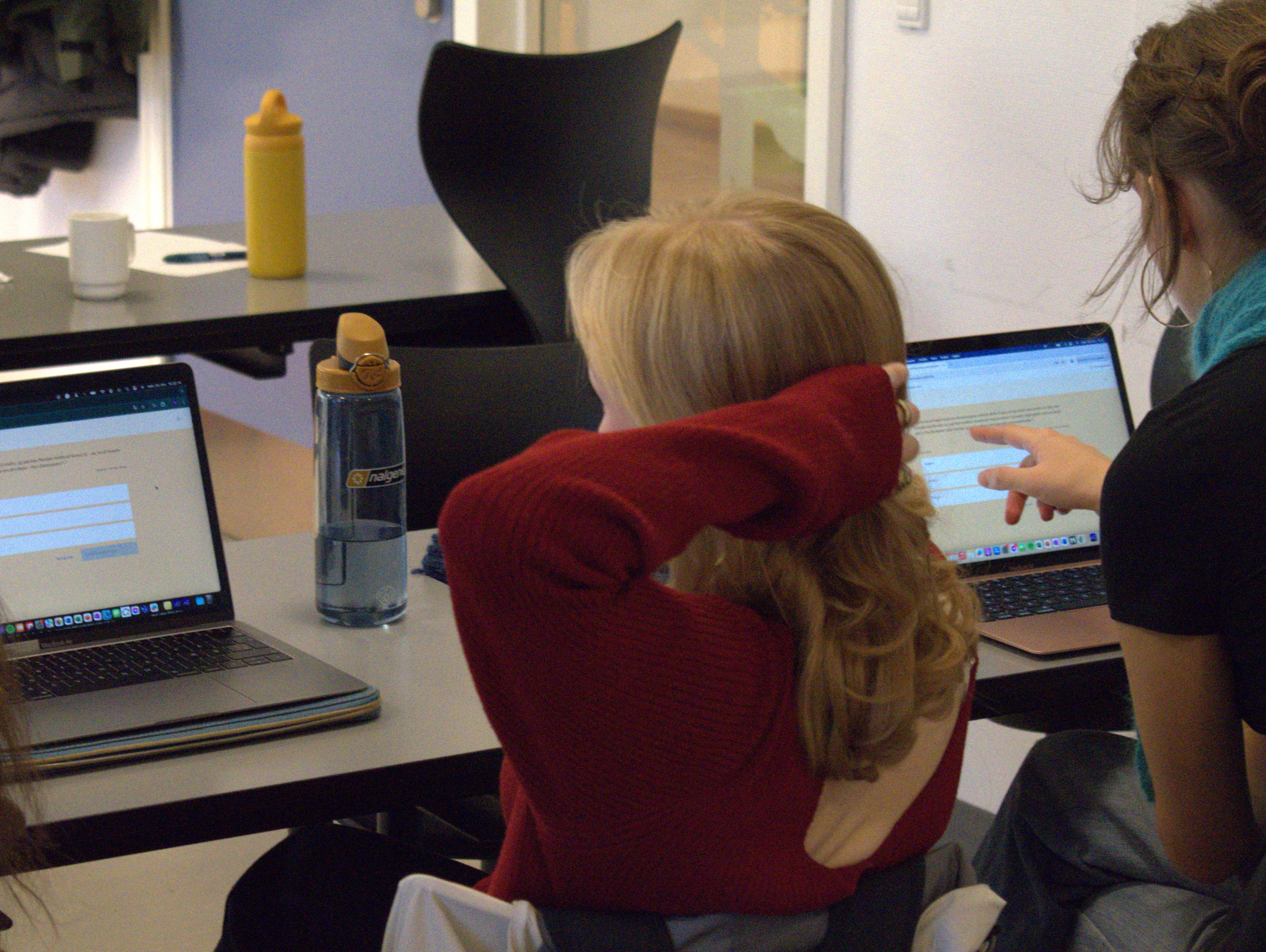}
    \end{subfigure}
    \begin{subfigure}[b]{0.32\textwidth}
        \includegraphics[width=\linewidth]{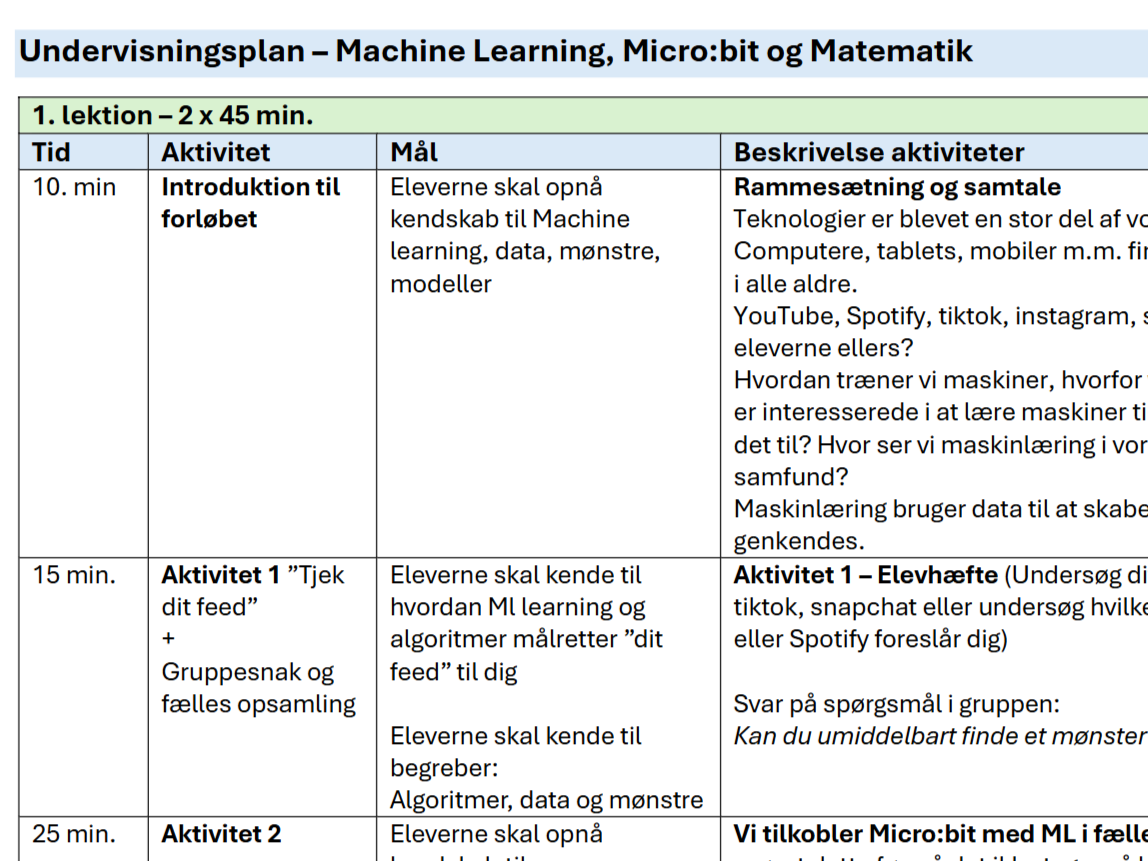}
    \end{subfigure}
  \caption{\textbf{AI in the classroom.}
  To support teachers and students in acting skillfully around AI, we explore how AI technologies and methods can interplay with knowledge and skills in existing school subjects.
  From left to right: Teachers engaging with AI activities as part of a workshop; students discussing sentences which they are annotating to train a language model; a teacher's lesson plan which integrates machine learning activities into the school subject. Figure used with permission from \citet{Bilstrup2025_AI_integration}.}
  \Description{Three images from the educational case study. Left: Teachers sitting around a table engaging with AI activities using paper slips as part of a workshop. Middle: students around a table discussing sentences which they are annotating on the computer to train a language model. Right: a table showing a teacher's lesson plan which integrates machine learning activities into the school subject.}
  \label{fig:education}
\end{figure*}

\subsubsection{Problem Domain}

AI tools are entering elementary and high school classrooms through commercially available, publicly accessible tools that every student and teacher can access.
These technologies have created a new reality in classrooms, where students are using AI tools (for good and bad), and teachers are being pushed to address how these technologies should play a role in the classroom~\citep{Berendt_AI_in_edu, Kasneci_chatgpt_for_good_education, okolo_AI_literacy_workshops}.
However, many teachers experience AI as a tool that students use to cheat and see their subjects changing beyond their control as a result of this development~\citep{Harvey2015_AI_harms_teachers}.

To address this, \citet{Bilstrup2025_AI_integration} explored how first language teachers can gain agency in how AI technologies and methods impact their subject teaching and empower them to integrate AI into their teaching practices and subject topics.
They specifically focused on how teachers and students can (a) use subject knowledge and skills to engage with AI and (b) use AI technologies and methods to become more skillful in the subjects.

\subsubsection{PAI Principles}

Following the principle of \mutualLearning, \citet{Bilstrup2025_AI_integration} put together a development team comprising high school teachers, educational researchers, NLP researchers, and digital humanities researchers.
This team developed educational tools and activities through workshops, prototyping, co-teaching sessions, and ongoing communication in a project spanning several months to explore how AI tools can be developed to fit existing classroom practices and how teachers can develop their practice to embrace the new AI technologies.
To enable teachers and students to reason about the opportunities and limitations of AI tools, they designed interactive tools that expose the inner workings of AI technologies.
For example, how computational tools can interpret words and sentences to provide new insights or representations of text pieces, and how this technology can be used to generate new text.
They supported these digital tools with unplugged activities~\citep{bell2018_CS_unplugged} where teachers and students can explore the data science practices behind AI systems without relying on external technologies they do not fully understand or control (see \citealt{Conelly2025_NLP_tools} for descriptions of the unplugged activities).

Through the workshops and design activities, they created a room where {\futureAlternatives} could be explored.
They identified first-language topics and learning goals that AI tools can be designed around, as well as opportunities for how AI and data science can provide new ways to engage with the subject content.
Here, digital humanities and NLP researchers provided examples of, and designed educational materials around, how AI technologies are used skillfully and methodically in contemporary literary research.  

Following the principle of \empowermentMediation, the project aimed to scale and sustain the research results through teacher engagement, the development of diverse learning materials around their AI tools, and by providing publicly available tools that work in classrooms.
To achieve this, \citet{Bilstrup2025_AI_integration} established a longitudinal partnership with one high school and its first-language teachers, where they co-designed classroom activities and new features in the AI tools, which they integrate into their regular teaching.
Based on these shared experiences, they developed workshops for first-language teachers at other schools, who can then remix the activities and integrate the tools into their own teaching practices.
This requires that teachers can count on the tools being available, robust enough for classroom use, and sustained over time.
Through these activities, they have begun to see a community of teachers who are uploading new datasets to the tools and sharing new ways to utilize them in various classrooms.
Furthermore, they presented experiences from the project to the association of first-language teachers for future collaboration.

\subsubsection{Design Challenges}

Addressing  \specialization, \citet{Bilstrup2025_AI_integration} explores how teachers and students can be empowered to fine-tune foundation models.
This occurs in classroom activities where students apply their subject knowledge to annotate text from the literature they are already familiar with.
Using one of the tools designed in the process described above, the teacher guides the classroom through the data science practice of pre-annotating, annotating, testing, and iterating on a language model.
The datasets are aggregated across different classrooms to train more powerful models on these larger datasets, which are made available to the contributing classrooms.
For example, students annotate the sentiment of sentences from H.C.\ Andersen's fairy tales.
This involves teaching of the languages and norms when fairy tales were written, close reading of each sentence, and discussions of different approaches to make the analysis.
Thus, they methodologically use their subject competencies to create a model with stronger capabilities in the topic they are teaching/learning about.

To address \emergentBehavior, they designed educational activities around the different steps towards developing advanced AI technologies which serve to engage teachers and students in reflections on how behavior can emerge from data and computational technologies.
This includes unplugged activities where teachers and students explore how text can be perceived as quantitative data, identify patterns in different datasets, and use digital tools to explore how these patterns can be used to generate new text.
Furthermore, they asked students to annotate the quality of arguments (from good to bad) for the purpose of engaging them in discussions on how norms are deliberately integrated into LLMs through manual labor. 

This also addresses the \humanAugmentation, as this provides an alternative to the zero-shot prompting usually conducted by teachers and students.
Instead of getting high-quality outputs, students take part in forming and discussing what dictates good and bad examples. 
Furthermore, this provides opportunities for discussing the methodology and comparing it with classical methodologies in the first language subject, thereby supporting students in reflecting on the possibilities and limitations of language models.

\subsubsection{Lessons Learned}

\citet{Bilstrup2025_AI_integration}'s project demonstrates the multidisciplinary competencies required to understand and leverage modern AI technologies, which often misalign with the siloed division of school subjects in the educational system.
A deeper understanding of the technologies and data practices behind AI was crucial for the teachers' ability to see AI as a tool they could take control of and use on their own terms.
At the same time, the first language teachers' knowledge of their subject was crucial for developing tools and activities where students' skills and knowledge from the subject came into play.
In the project, AI became a driver for the development of school subjects and teaching practices among teachers.
The tools and activities also engaged students in using school subject competencies to contest AI outputs and reflect on their capabilities.

%% file: content/04.2-case-creativity.tex
\subsection{Case Study 2: Creativity}
\label{sec:creativity}

Generative AI reshapes creative work both concretely, as a new means of producing text, images, and audio, and structurally, by reorganizing how creative work is produced, evaluated, and paid for across platforms and tool chains.
This case draws on two initiatives: interviews and participatory engagements with 30 professionals in graphic design and music production, and the design and evaluation of a research prototype for AI-supported design ideation.

Interviewees reported learning system behavior through trial, error, and informal exchange rather than any dialogue with developers; the \mutualLearning{} gap is structural, with creators' domain knowledge never reaching model developers and technical knowledge never reaching practitioners.
Generative tools also fit poorly into existing assemblages of software, formats, and institutional routines, pulling work into proprietary environments organized around prompting, curation, and selection~\citep{Chang2023} (\artifactEcologies).
Participatory discussion reframed AI as an instrument that mediates activity rather than dictates outcomes, with empowerment located in creators' ability to decide which parts of their practice are open to delegation (\empowermentMediation).

Foundation models are adapted to creative domains without the practitioners whose standards they affect, so \specialization{} proceeds without access to the tacit criteria of professional quality, and shared datasets and defaults push outputs toward stylistic sameness~\citep{Anderson2024, Dalsgaard2025}.
The prototype evaluation supplied the case's central mechanism for \humanAugmentation: pairs of designers welcomed AI-generated concepts during fast divergent ideation but consistently set them aside in convergent phases, relying on situational knowledge and personal style.
That boundary was not built into the interface; it emerged through use.
Authorship and value distribution, with recognition and compensation flowing toward platform owners~\citep{Samuelson2023, Van_Dijck2018}, surfaced as questions of governance rather than design.

Participation must reach into specialization and boundary-setting, and augmentation is negotiated in practice rather than designed in advance.
Sustaining either requires collective structures, professional associations and unions among them, which individual projects cannot supply.
The case is diagnostic rather than interventional, participation was uneven across freelancers, studios, and marginalized groups, and longitudinal effects on portfolios and income remain unmeasured.

Generative AI is rapidly reshaping the creative industries: \textbf{concretely}, by offering new means of generating content such as text, visuals, and audio; and \textbf{structurally}, by reorganizing how creative work is produced, evaluated, and economically valued across platform ecosystems and professional tool chains.
In this case study, we examine how generative AI reshapes creative work by drawing on two research initiatives:
first, a study consisting of interviews and participatory engagements with 30 professionals in graphic design and music production; and second, the development and evaluation of a research prototype for AI-supported design ideation.
Through these, we address how AI enters everyday creative workflows, how it reconfigures professional norms and economic conditions, and how questions of authorship, skill, and creative identity are negotiated when AI systems become part of the production process.

\begin{figure}
    \centering
    \includegraphics[width=0.75\linewidth]{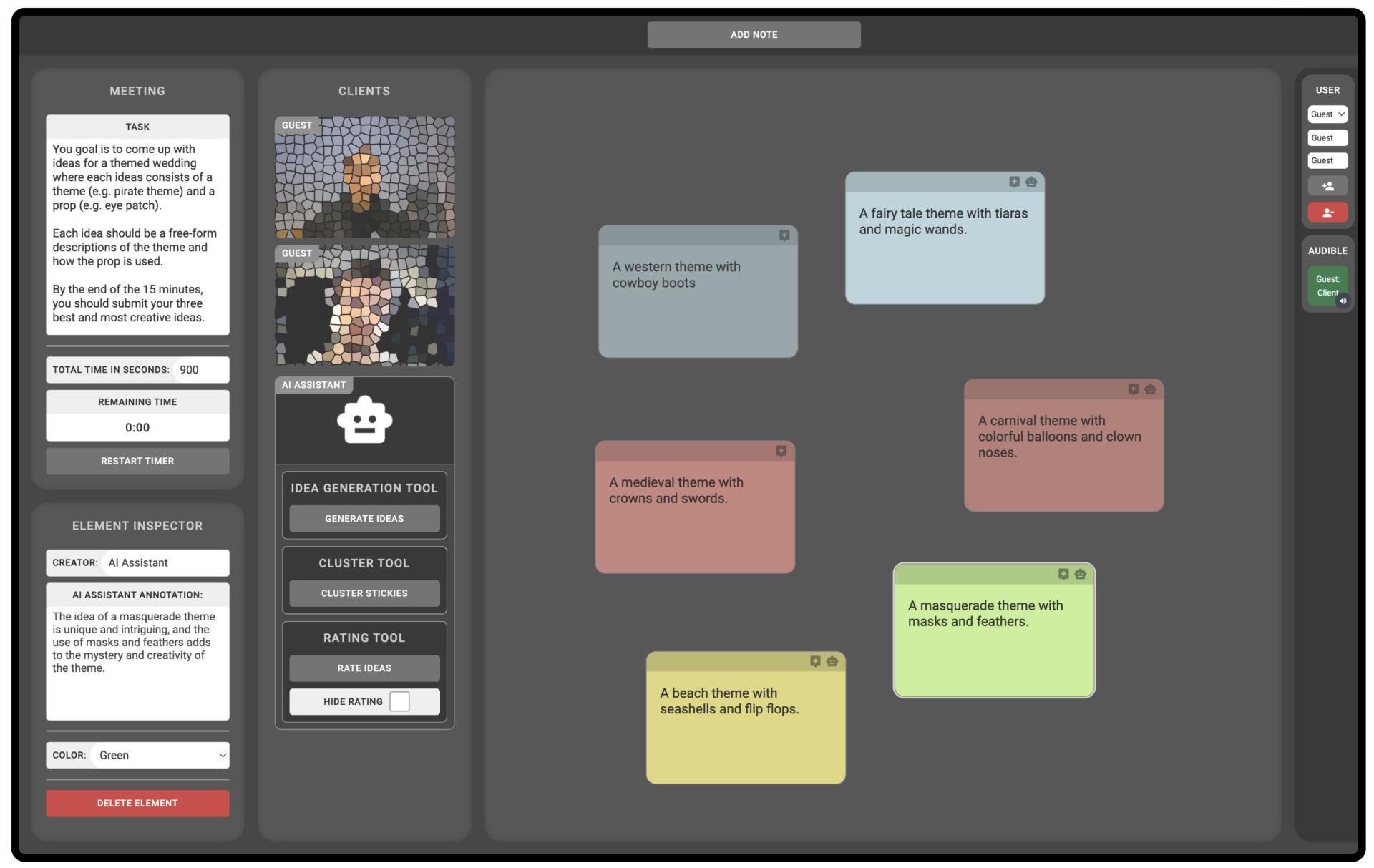}
    \caption{\textbf{Using GenAI to support creativity.}
    A research prototype using GenAI to support divergent and convergent phases of team-based ideation.}
    \label{fig:creativity}
    \Description{A screenshot from a prototype system for ideation supported by generative AI. The main interface is a series of colored digital sticky-notes each representing an idea. A control panel lets users ask the AI to generate a new idea, cluster existing ideas, or rate the existing ideas. Neither has yet been done by the participants.}
\end{figure}

\subsubsection{Problem Domain}

Generative AI is increasingly integrated into creative work through widely available platforms and professional tool chains.
Systems such as ChatGPT, Midjourney, Adobe Firefly, and platform-embedded assistants are now used for ideation, variation, refinement, and production across creative domains.
For creative professionals, this shift raises both practical and normative concerns.
On a practical level, AI alters workflows by accelerating some phases of work while compressing others, often privileging speed and volume over deliberation and revision.
On a normative level, it unsettles how quality is assessed, how authorship is attributed, and how expertise is recognized.
These concerns extend beyond individual practices to the collective standards maintained by professions, communities, and institutions within the creative industries.
The problem, therefore, is not simply whether generative AI can produce acceptable outputs, but how its integration reshapes the conditions under which creative work is performed, evaluated, and sustained.

\subsubsection{PAI Principles}

In creative work, \mutualLearning becomes more difficult but also increasingly necessary with the current generation of off-the-shelf AI services.
As we have learned from interviews with creative practitioners across domains of art, graphic design, and music production, generative AI systems are typically introduced with limited transparency regarding how outputs are produced or how models have been trained.
As a result, creators often learn about system behavior through trial, error, and informal exchange rather than through dialogue with designers or developers.
This places the burden of interpretation on practitioners, who must adapt their practices to system assumptions that may not align with established notions of quality, originality, or authorship, in stark contrast with the PD principle of designing tools for skilled users rather than automated systems.
The mutual learning gap here is structural: creators hold deep domain knowledge that never reaches model developers, while developers hold technical knowledge about system behavior that never reaches practitioners.
Bridging this gap requires sustained engagement between these groups, which is precisely the kind of collaborative knowledge exchange that PD has long advocated for.

The principle of \artifactEcologies is particularly salient in creative domains.
Creative work is already structured around complex assemblages of software, file formats, hardware, and institutional infrastructures.
Generative AI does not enter a neutral environment but must be negotiated into these existing ecologies.
Participants described how AI tools often failed to integrate cleanly with established workflows, pushing work into platform-specific environments and shifting creative activity toward prompting, curation, and selection within proprietary systems~\citep{Chang2023}.
As creative tasks are increasingly automated and absorbed into platform environments, practitioners face challenges similar to those addressed in workplace contexts in the early days of participatory design (see Section~\ref{sec:history}).
This resembles the distinction between the tool perspective and the system perspective introduced in PD in the early 1980s.

Concerns around \empowermentMediation were prominent in our interview data.
Many participants described how current AI platforms emphasize throughput and scale rather than care, judgment, or craft.
While prompting and curating AI outputs can itself involve considerable skill and judgment, practitioners worried that platform defaults compress the range of recognized expertise, reducing multifaceted creative processes to a narrower set of interactions with generative systems.
Participatory discussions surfaced alternative framings in which AI was understood as an instrument that mediates activity rather than dictates outcomes.
In this view, empowerment lies in preserving creators' ability to define which aspects of their practice are open to delegation and which remain central to professional identity.

\subsubsection{Design Challenges}

The challenge of \specialization is acute in creative work.
Foundation models are adapted to creative domains with minimal involvement from the practitioners whose traditions and standards they affect.
Interviewed creators emphasized that these specialization decisions are made without access to the tacit criteria that govern professional quality. 
This means that outputs may appear competent or stylistically pleasing yet fail to capture domain-specific nuances that matter in practice.
These gaps carry broader consequences: common datasets and shared defaults can reinforce stylistic sameness, a tendency described as \textit{generative flattening} or \textit{homogenization}~\citep{Anderson2024, Dalsgaard2025}.
Medium-specific craft gives way to fluency in AI services, with the risk that traditional expertise is devalued and aesthetic outcomes converge on a narrower range~\citep{Oppenlaender2025, Huang2024}.
A participatory approach to specialization could counter these tendencies through datasets co-curated with practitioners, iterative feedback cycles where creators flag divergence from domain standards, and plural specializations that allow several creative directions to coexist.

A PD perspective treats \humanAugmentation differently: the goal is not to automate whatever can be automated, but to extend human agency while preserving the practices that creators deem central to their identity.
This means asking which aspects of creative work are open to acceleration---routine drafting, stylistic variation, bulk refinement---and which are non-negotiable, such as the interpretive judgment of an editor or the personal style of a composer.
We observed this directly through the development and evaluation of a research prototype for AI-supported design ideation, built through iterative participatory engagement with professional designers.
The prototype supports collaborative design sessions by generating visual concepts in response to design briefs.
Evaluating it with pairs of designers (see Figure~\ref{fig:creativity}), we found a clear boundary in how practitioners negotiated AI involvement: AI-generated concepts were welcomed during fast-paced ideation, where volume and variety supported exploration, but in later phases designers consistently preferred to work without AI suggestions, drawing instead on situational knowledge and domain judgment.
This boundary was not predetermined but emerged through use, illustrating how augmentation is negotiated in practice rather than designed in advance.

Questions of authorship and value distribution surfaced repeatedly across interviews.
When AI-generated outputs draw on uncredited training data (or even illegally stolen copyrighted artifacts), attribution becomes ambiguous, and recognition and compensation tend to flow toward platform owners and model providers rather than individual practitioners~\citep{Samuelson2023, Van_Dijck2018}.
For many interviewees, this undermines both individual creative identity and the collective capacity of creative communities to set and defend their own standards; questions not of interface design but of governance.

\subsubsection{Lessons Learned}

Viewing creative work through a participatory AI lens highlights how generative systems intervene not only in workflows but in the normative foundations of creative practice.
Across interviews and prototype evaluations, creators consistently framed generative AI as a force that reorganizes where judgment, authorship, and responsibility are located.
These shifts are understood less as technical changes than as transformations in professional identity and collective standards.

A central lesson is that participation must extend into processes of specialization and boundary-setting.
When creators were involved in articulating which aspects of their work could be delegated to AI and which should remain under human control, AI shifted from being perceived as an external imposition to a negotiable instrument.
The prototype evaluations made this concrete: designers drew clear lines between phases where AI exploration was productive and phases where it was unwelcome, and these lines emerged through participatory use rather than prior specification.

A second lesson is that augmentation is a contested achievement rather than an inherent property.
Augmentation was welcomed when AI supported exploration or variation, but resisted when it compressed interpretive judgment or obscured authorship.
Designing for human augmentation therefore requires participatory negotiation of pace, visibility, and control, not only interface design.

At a collective level, the case underscores the importance of \textbf{participatory governance}.
Participants emphasized the limits of individual action: professional associations, unions, and informal creative communities were identified as necessary mediators for sustaining participatory influence beyond isolated projects.
Without such collective structures, participatory gains remain fragile and dependent on the goodwill of platform providers.

Despite these insights, notable limitations remain.
Longitudinal studies are needed to reveal how participatory efforts actually affect creative portfolios, recognition, and economic outcomes over time.
Participation itself is uneven: freelancers, smaller studios, and marginalized groups are often underrepresented, risking results that favor those already well-resourced.
And even when participatory mechanisms are well-designed, infrastructure can be fragile, such as when provenance and credit signals get lost as work moves through different tools and platforms.

%% file: content/04.3-case-online-knowledge.tex
\subsection{Case Study 3: Online Knowledge}
\label{sec:online-knowledge}

Wikipedia is maintained almost entirely by volunteers~\citep{CollWiki}, and the maintenance is not keeping up: millions of orphan articles have no incoming links and reach few readers~\citep{orphans, wikipedia_orphan}, while existing pages need constant updating, referencing, and linking~\citep{add_link, entity_insertion}.
These tasks demand contextual judgment and fluency in Wikipedia's norms of neutrality and verifiability.
The Wikimedia Foundation's humans-first AI strategy~\citep{wmf_humans_first} sets the constraint: AI as scaffolding for editors, never a replacement.

\winsert~\citep{yildirim2025wikinsert}, built on the multilingual \xloki framework~\citep{entity_insertion}, scores the relevance of concepts to sentences in an article and overlays a heatmap on live Wikipedia pages through a browser extension.
It edits nothing; it visualizes cues, and editors decide whether to rewrite, link, or ignore (\empowermentMediation).
The overlay design fits Wikipedia's existing ecology of bots, templates, and editing tools rather than adding a parallel interface (\artifactEcologies), and the dialogue between editors and researchers, in which editors learned what the model could and could not do while researchers adapted it to community norms, made augmentation an explicit choice over an automation path that was open early on (\mutualLearning, \futureAlternatives).
Surfacing orphan and underrepresented content aligns the tool with the movement's knowledge-equity mission, and communities keep final say over inclusion (\emancipatoryDemocracy).

The central mechanism is \humanAugmentation{} as a negotiated outcome: a preliminary study with five editors involved in Wikipedia AI development found that they folded the tool's suggestions into their own strategies and built working mental models without full algorithmic detail, a property the study calls functional transparency.
Irrelevant or biased suggestions were unavoidable (\emergentBehavior), so editors also helped define what a good recommendation is; adapting general models to editorial norms, such as verifiability for citation suggestions, required \specialization{} guided by editor feedback; and the platform's 300-plus language editions and mixed content types mark \multimodality{} as the next frontier.

Humans-first AI is workable, interpretable cues serve trust better than full transparency, and participation is infrastructural rather than optional.
The evaluation is small, and extending participatory governance across hundreds of language communities remains open.

Wikipedia provides a striking case study to understand how PAI principles may be applied in long-term processes of use.
It is one of the largest and most influential online knowledge platforms, maintained almost entirely by volunteers~\citep{CollWiki}.
Yet sustaining it is increasingly difficult.
Millions of orphan articles~\citep{wikipedia_orphan} remain unlinked and invisible to readers~\citep{orphans}, while existing pages demand constant maintenance to stay up-to-date~\citep{add_link, entity_insertion}.
Editors face demanding tasks of organizing, adding, updating, and linking knowledge to keep the encyclopedia coherent~\citep{BecomingWiki, CareWiki}. 
AI can help---but only if designed in line with Wikipedia's humans-first AI strategy~\citep{wmf_humans_first}, which positions AI as scaffolding for volunteers rather than a replacement.
This case study unfolds the possibilities of offering PD's principles to such use situations.

\begin{figure}
    \centering
    \includegraphics[width=\linewidth]{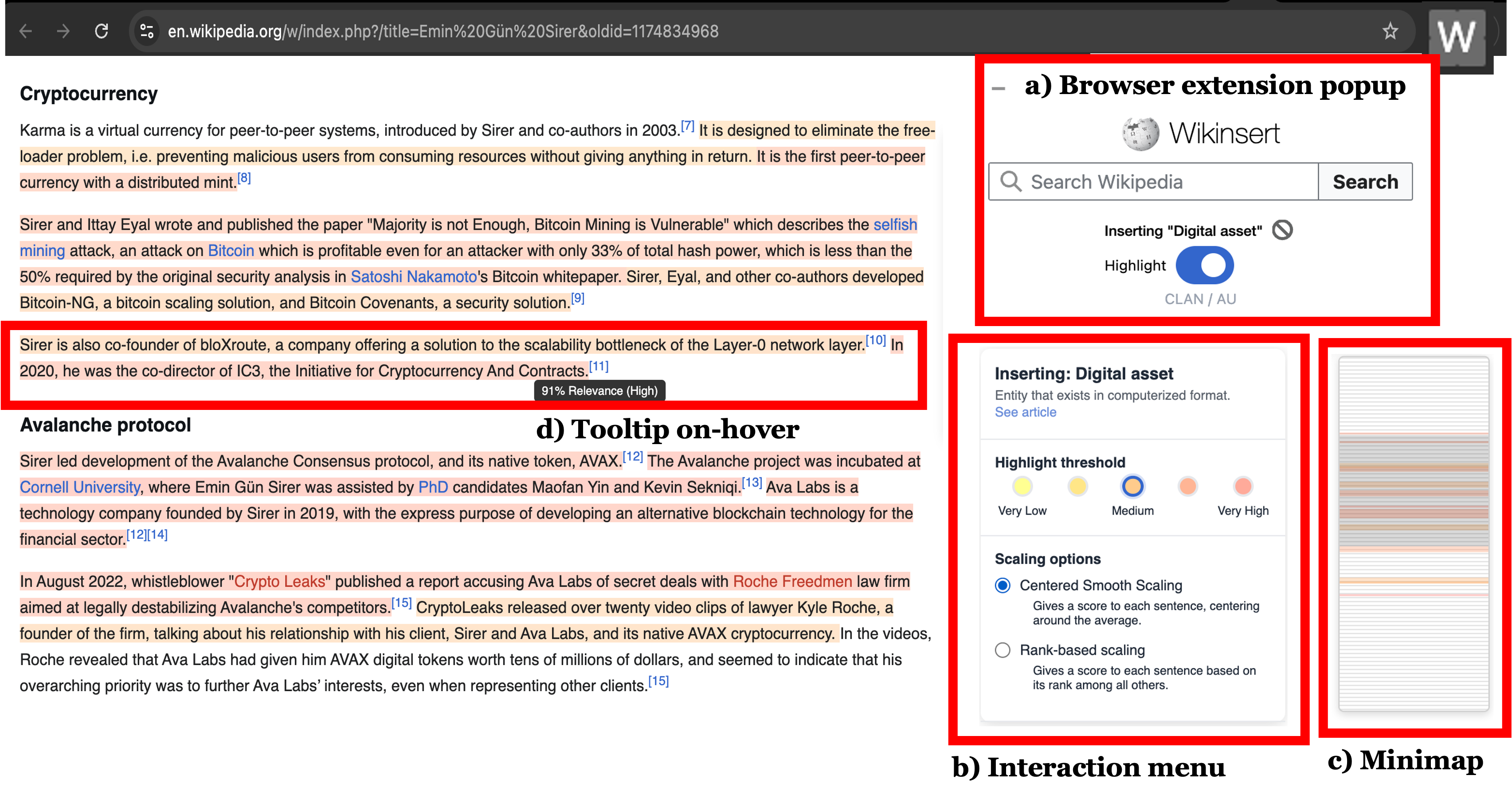}
    \caption{\textbf{Human-in-the-loop knowledge maintenance.}
    An excerpt from the Wikipedia article ``\texttt{Emin Gün Sirer}'' showcasing \winsert's entity-aware highlighting via an overlaid heatmap (redder regions indicate higher relevance, whereas whiter regions indicate lower relevance) and the interface components: minimap for global navigation and interaction menu panel for threshold adjustment. 
    }
    \label{fig:knowledge}
    \Description{%
    Screenshot of a Wikipedia page for Emin Gün Sirer with \texttt{Wikinsert} browser extension overlays. Text passages are highlighted in a red–white heatmap (redder text indicates higher entity relevance). Four interface components are shown: (a) a browser extension popup for search and highlighting control, (b) an interaction menu with options for highlight threshold and scaling methods, (c) a minimap visualization of highlights across the whole page, and (d) an on-hover tooltip displaying entity relevance for a selected sentence. Layout is rearranged for demonstration purposes.
    }
\end{figure}

\subsubsection{Problem Domain}

Knowledge gaps~\citep{knowlege_gaps_zia} in Wikipedia manifest in two interrelated ways~\citep{knowledge_gaps_taxonomy}. 
Content gaps occur when topics are missing or underrepresented: for example, biographies of women or Global South perspectives~\citep{wagner2015man}.
Structural gaps occur when articles exist but remain isolated from the larger knowledge graph.
For example, orphan articles with no incoming links, are effectively invisible unless readers know exactly what to search~\citep{orphans}.

Research on knowledge navigation confirms the impact of these gaps~\citep{akhil_phd_thesis}. 
Orphan articles consistently attract fewer readers, and de-orphanization (connecting them to other pages) measurably increases visibility~\citep{orphans}. 
Beyond orphans, editors must also perform constant maintenance: updating pages when new concepts emerge, adding references, and linking entities so that readers can move seamlessly between related topics~\citep{add_link, entity_insertion}.
These are demanding editorial tasks, requiring both contextual judgment and deep familiarity with Wikipedia's norms of neutrality and verifiability.

\paragraph{\winsert: Human-in-the-Loop Knowledge Maintenance}

\winsert was designed to support editors in maintaining Wikipedia's knowledge network~\citep{yildirim2025wikinsert}.
Built on the multilingual \xloki framework~\citep{entity_insertion}, it computes relevance scores between concepts and sentences in source articles, then overlays a heatmap directly onto live Wikipedia pages through a browser extension. 
Sentences predicted as good candidates for adding or updating a link are highlighted in color, providing editors with at-a-glance scaffolding for where their attention might be most valuable (cf. Figure~\ref{fig:knowledge}).

The architecture is deliberately lightweight: heavy machine learning computations happen offline, and the browser extension only fetches precomputed scores from a backend API.
The tool does not \emph{edit autonomously; it only visualizes cues}. 
Editors remain in full control of deciding whether to rewrite text, add a link, or ignore the suggestion entirely.
This reflects a ``humans-first'' design ethos: AI provides transparent signals while leaving authority with the community.

A preliminary user study involving five Wikipedia editors who have also been involved in AI development on Wikipedia confirmed that editors used \winsert in interpretive ways. 
Participants combined AI suggestions with their own strategies, valuing the tool for reducing search time while maintaining judgment over final edits.
Importantly, they developed effective mental models of the system without needing full algorithmic transparency; what the study terms \textit{functional transparency}. 
This demonstrates how well-designed scaffolds can enhance fluency without displacing expertise.

\subsubsection{PAI Principles}

\winsert embodies \empowermentMediation by strengthening editors' collective ability to keep knowledge current.
Rather than automating edits, it mediates between vast sources of candidate knowledge and the editor's limited attention, reducing cognitive load while amplifying human agency.
This aligns with PD's goal of building community capacity, not individual dependency.

Knowledge gaps often reflect systemic inequities.
By surfacing opportunities to link orphan articles or integrate underrepresented topics, \winsert contributes to the Wikimedia movement's mission of knowledge equity (\emancipatoryDemocracy).
Its multilingual capabilities extend this impact across smaller and marginalized language editions. 
Democracy is preserved: the tool offers suggestions, but communities retain final authority over what is included.

Designing \winsert required spaces for editors and researchers to learn from one another (\mutualLearning).
Editors gained literacy in what AI could (and could not) do, while researchers adapted models to align with Wikipedia's norms.
This reciprocal process reflects PD's vision of mutual learning as a sustained, iterative practice.

PD emphasizes that technological trajectories are not predetermined.
Early design choices could have prioritized automation, but participatory dialogue surfaced an alternative: augmentation.
By openly considering multiple \futureAlternatives, the community chose alignment with Wikimedia's humans-first ethos.

Wikipedia is embedded in a dense ecology of \artifactEcologies: bots, templates, editing tools, citation guidelines.
\winsert was designed to integrate seamlessly into this environment, using a browser extension overlay rather than creating a parallel interface.
Recognizing these artifact ecologies ensured adoption by aligning with existing practices.

\subsubsection{Design Challenges}

Perhaps the most central challenge was \emph{ensuring augmentation rather than automation} (\humanAugmentation).
Editors remained co-creators of system behavior: their inputs, feedback, and selective uptake shaped how the model's suggestions were interpreted.
This reframed interaction design as a participatory data practice, grounding augmentation in agency and context.

AI recommendations are probabilistic, and emergent behaviors (irrelevant suggestions, reinforcement of biases) were unavoidable (\emergentBehavior).
Participatory feedback loops became essential: editors helped define what counted as a ``good recommendation,'' not only technically but socially.
This highlighted the need for continuous renegotiation of system behavior over time.

Adapting general language models to Wikipedia's editorial norms required careful attention to\\ \specialization.
For instance, citation suggestions had to respect verifiability rules.
Editor feedback was critical to aligning the system with community expectations.

Wikipedia spans over 300 languages and multiple modalities (text, images, infoboxes; an example of the need for \multimodality).
\winsert focused initially on text, but its multilingual backbone illustrates the need for distributed, inclusive design. 
Transparency about how data is used helps editors make informed choices about adoption.

\subsubsection{Lessons Learned}

This case demonstrates how PD can guide the design of AI for online knowledge.
Several lessons emerge:

\begin{itemize}
    \item \textbf{Humans-first AI is viable:} \winsert shows that AI can enhance knowledge work without displacing it.
    \item \textbf{Interpretability matters more than transparency:} Simple, actionable cues support trust better than full algorithmic detail.
    \item \textbf{Participation is ongoing:} Feedback and adaptation are infrastructural, not optional.
    \item \textbf{Equity can be operationalized:} Targeting orphan and underrepresented content aligns AI with emancipatory goals.
    \item \textbf{Sociotechnical fit is critical:} Integration into existing workflows enabled adoption.
    \item \textbf{Scalability is the next challenge:} Extending participatory AI across global communities requires new governance mechanisms.
\end{itemize}

Taken together, the case affirms that AI for online knowledge should function as scaffolding for human judgment, not a substitute for it.
By grounding design in PD principles and facing the challenges of augmentation, emergence, specialization, and multimodality, tools such as \winsert illustrate how communities can remain stewards of their knowledge infrastructures, even in an AI-driven future.

%% file: content/04.4-case-agriculture.tex
\subsection{Case Study 4: Agriculture}
\label{sec:agriculture}

Weather forecasts govern when farmers plant, spray, and harvest, and machine learning now delivers hyper-local predictions~\citep{DBLP:journals/corr/abs-2410-08641}.
Their value, however, depends on how farmers read them and on what they do to established practice.
This case follows a startup that sells such predictions to farmers, fed by weather stations placed in the farmers' own fields.

Farming decisions rest on a mix of personal measurements (rain gauges at chosen spots, visual inspection of soil and crops), public forecasts, and inherited knowledge, an ecology the company had to learn before its product could fit it (\artifactEcologies).
It ran workshops with farmers on prototypes of the weather station and the application, and seated a local farmer on its advisory board to shape product and strategy over time (\mutualLearning, \futureAlternatives).
Farmers site and maintain the stations themselves, which lets them tune the spatial granularity of the forecasts they receive; some now dispatch rented combine harvesters field by field on the model's local predictions instead of pre-planning on public forecasts (\empowermentMediation).

The central mechanism is a negotiation over how model output is shown.
The model expresses its uncertainty as probabilistic maps, which are more accurate than the sharp radar images farmers expect and which farmers read as worse (\textit{``this looks different, it's probably worse''}); the company redesigned the maps toward a middle ground that stays faithful to the model, and some farmers in turn began to question their own practices.
Other customers, such as racing teams wanting minute-level forecasts three hours out rather than farmers' 8--12 hour horizon, raise \specialization{} as a choice between one general model and a fractured set of specialized ones.
Under \multimodality, the company owns both the stations and the data they produce: farmers guard field data because it affects prices and competition, and accept the arrangement in exchange for better predictions, which cuts against the data-ownership stance we otherwise favor~\citep{DBLP:conf/fat/TsengYQRS25}.

The company credits participation with correcting an accuracy-only view of what matters and with opening new customer groups, and it wants a lighter, more agile form of PAI compatible with time-to-market pressure.
The benefits nonetheless accrue to larger farms with the resources to run data infrastructure, and farmers do not own their data.

This case study examines how a startup company developed farmer-centered AI weather prediction tools, exploring how participatory design principles can ensure that advanced forecasting technology serves agricultural communities rather than disrupting established knowledge and workflows.

\begin{figure}
    \centering
    \includegraphics[width=0.8\linewidth]{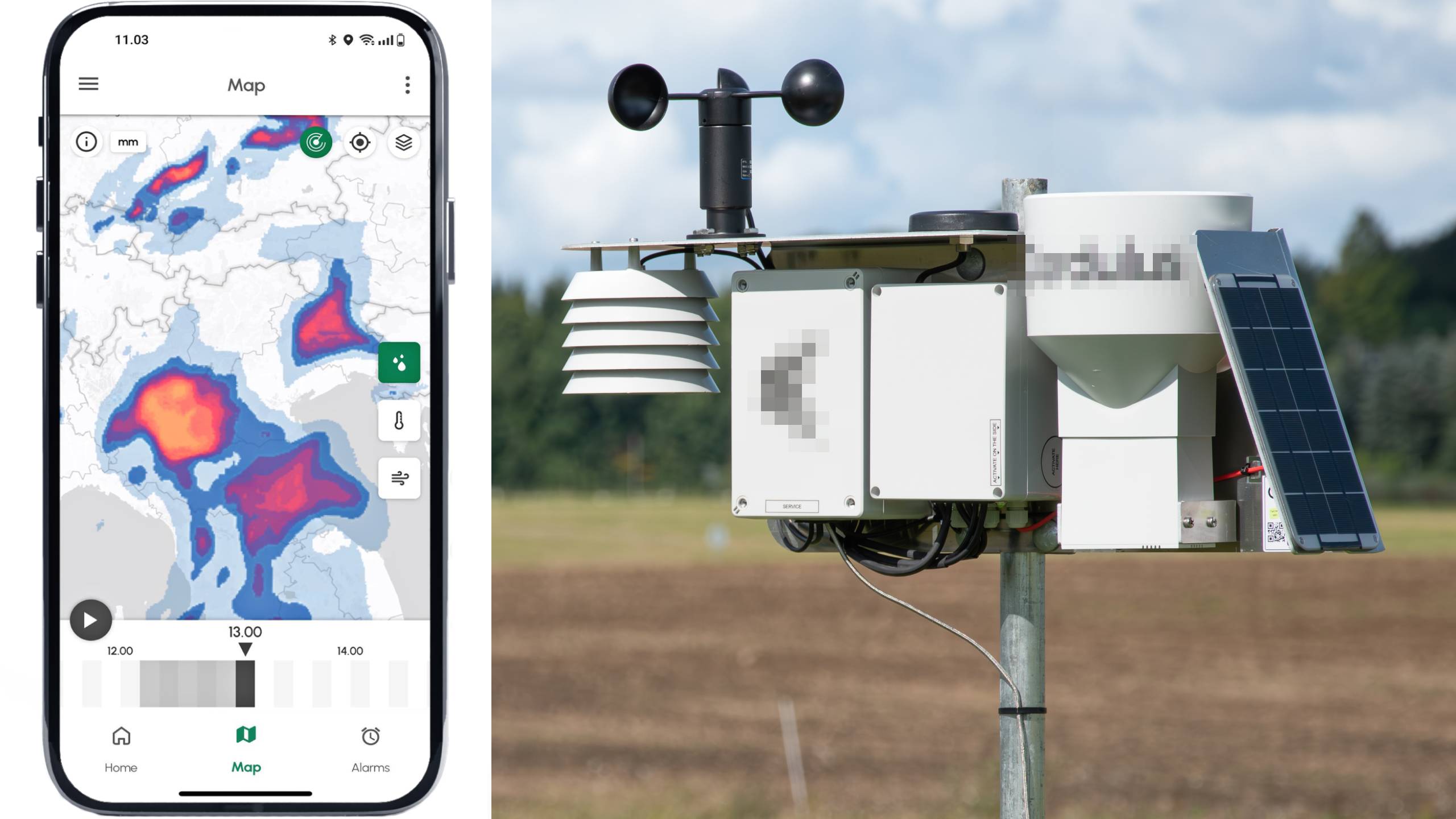}
    \caption{\textbf{Farmer-friendly AI weather predictions.}
    (Left)~Using machine-learning-based models, the startup is able to provide hyper-local weather predictions to farmers via a mobile application. 
    (Right)~Physical weather stations in the fields automatically collect data to train and inform the prediction models.}
    \label{fig:cordulus}
    \Description{%
    On the left is a smartphone screen displaying a weather map. The map shows precipitation patterns in different colors, with red and purple areas indicating heavy rainfall and lighter blue areas showing lower rainfall. A time slider at the bottom allows the user to check forecasts at different hours, and icons on the right provide options such as temperature and other weather data layers.
    On the right is a weather station installed outdoors in a field. The station includes various instruments: an anemometer with rotating cups on top for measuring wind speed, a white radiation shield with slats for housing temperature and humidity sensors, and a large white rain gauge. A solar panel is mounted on the side to power the device. The equipment is attached to a metal pole, with visible cables running underneath.
    }
\end{figure}

\subsubsection{Problem Domain}

Weather prediction is an essential factor in agriculture, informing farmers when and where to plant and harvest crops, or spray pesticides and fertilizers. 
Recently, machine learning-based models have enabled hyper-local weather predictions.
Beyond predictive accuracy, the value of such forecasts depends on how they are delivered and understood by farmers, and on their implications for agricultural practices.
We explore how a startup company provides farmers with hyper-local weather predictions and discuss how PAI design principles apply.

\subsubsection{PAI Principles}

Farming is rich in traditions and communal practices.
This extends to the decision-making about when, where, and what to grow and harvest.
For example, farmers often place rain gauges at key points in their fields and visually inspect plants and soil to determine how best to fertilize, apply pesticides, or deploy harvesters.
In making these decisions, they draw on their own measurements and publicly available weather data, but also on traditional knowledge. 

When the startup company introduced an AI model for hyper-local weather predictions to the farming market~\citep{DBLP:journals/corr/abs-2410-08641}, they had to engage with farmers to learn about these \artifactEcologies{} of tools, traditions, and practices surrounding their product.
Following the principle of \mutualLearning{}, the company invited farmers to workshops where prototypes of their weather stations and user-facing application design were discussed (Figure~\ref{fig:cordulus}).
They further included a local farmer on their advisory board to continuously guide the product design and strategic decision-making, fitting with the principle of \futureAlternatives{}.

Throughout this process, they uncovered assumptions about agricultural practices built into their product and service design which conflicted with the conventions, tools, and traditions.
For example, farmers are used to seeing sharp, discrete weather maps.
The startup's weather maps, by contrast, make the uncertainty of their predictions transparent, outputting probabilities.
While these predictions were more accurate, they conflicted with the established practices (\textit{``You are used to seeing a weather radar [\ldots] that looks sharp''}, \textit{``[the] end-user might say: `this looks different, it's probably worse.'\,''}).
This motivated the company to change the visual design to a middle ground that is closer to farmers' expectations, while still faithful to the AI model output. 
At the same time, some farmers started to question their practices to benefit from the superior accuracy of data-driven decision-making.

The rise of AI technology also falls into a broader trend affecting the agriculture community.
As regulatory demands and market pressure grow, smaller farms are struggling to compete and consolidate into larger organizations with more employees and acreage.
These larger organizations are better equipped to dedicate resources to data collection and analysis and to the use of AI.
For farmers using the more accurate and fine-grained weather predictions, new practices are starting to emerge.
For example, during harvest, it is common to rent expensive machinery, such as combine harvesters.
It is essential for the farmers to use these resources as effectively as possible.
Using the more accurate and fine-grained predictions by the startup's weather model, farmers are beginning to change how they allocate this machinery.
Instead of basing routes on public weather forecasts and adjusting as they change, some have begun to use the model's local predictions to continuously send harvesters to those fields that offer the best conditions.
By placing the weather stations at the most critical sites, the farmers can influence the granularity profile of these weather predictions, further strengthening their basis of decision-making and contributing to their \empowermentMediation{}. 

\subsubsection{Design Challenges}

The startup's AI model was originally developed to suit farmers.
Yet, it quickly became apparent that other stakeholders can equally benefit from fine-grained predictions (\specialization).
Stakeholders differ in what they expect from the model.
For example, while farmers look for predictions between 8--12 hours ahead of time, other stakeholders, such as racing teams, require a minute-by-minute prediction for the next three hours.
Much of this divergence can be addressed through adaptations of the user interface, but it also provides opportunities to specialize the underlying model. 
Here, the challenge lies in determining the trade-offs between training and maintaining a fractured ecosystem of specialized models, or using more general models with potentially reduced fit to different user groups.

As for \multimodality{}, the startup's prediction model relies on sensor inputs supplied by local weather stations.
While farmers have autonomy over the placement and maintenance of these weather stations, both hardware and produced data are owned by the startup company. 
This was a necessary choice, as farmers are used to owning equipment, but are highly sensitive to exposing data about their fields and operations publicly, since this can directly impact prices and competition.
In return for sharing their data directly via weather stations with the startup, farmers receive higher-quality predictions.
While we generally consider data ownership and sharing as beneficial for user-driven processes~\citep{DBLP:conf/fat/TsengYQRS25}, this case highlights the need for approaches that both preserve data privacy, but still offer sufficient value to support the development of new AI models. 

\subsubsection{Lessons Learned}

There are several observations that can be made from looking at AI and Participatory Design from the standpoint of a startup company working with a specialized customer group, such as the agriculture community.
Importantly, the startup company acknowledges that PAI is a useful counterweight to overly focusing on incremental improvements in traditional AI measures such as accuracy alone, and to identify what really matters to stakeholders.
As with other user-centered practices, it can be challenging to apply PAI principles in a competitive market, particularly as a startup company that needs to establish a successful business model.
However, this case study indicates that PAI can lead to better products (improved map visualizations) and allow a company to better cater its product to new customer groups (racing teams).

We thus see a strong potential for PAI to help resolve the tension between time (to market) and product/service quality by strategically involving stakeholders.
We believe that, to be successful in competitive environments, PD must be applied holistically to the entire process: from traditional PD for the development of hardware and software to PAI for training and usage of models.
Yet, the feedback and discussions indicate that, for practical use in a time and money-sensitive environment, a more agile---and thus flexible---version of PAI could allow for PAI benefits in otherwise constrained application cases.

%% file: content/04.5-case-manufacturing.tex
\subsection{Case Study 5: Manufacturing}
\label{sec:manufacturing}

Industrial quality assurance (QA) asks workers to measure parts and catch production errors quickly, under frequent job rotation that leaves little time to learn each task and with routine work that drives turnover~\citep{BeloFenderFeuchtner2019, BeloWissingFeuchtner2023}.
This case follows a project in which LEGO Group QA workers took part in designing an augmented-reality measurement aid driven by a CNN that recognizes industrial objects and their pose from high-resolution video.

Workplace visits, interviews, and brainstorming in the manner of future workshops~\citep{Boedker1995, FutureWorkshops} preceded any technology choice; the idea of AI-guided measurement came out of these sessions, as did the finding that near-symmetrical objects were the main source of error (\mutualLearning).
Designers then demonstrated AR glasses, projection, and display alternatives.
Workers rejected the glasses and steered the design toward a display integrated into the table surface, and a second prototype simplified the setup to a single phone camera on a stand~\citep{BeloWissingFeuchtner2023} (\futureAlternatives).
Any solution also had to work with the CAD system supplying brick drawings and the database holding quality assessments (\artifactEcologies).
When COVID-19 lockdowns interrupted the work, the team built a remote cooperative-prototyping tool to keep the sessions going.

Placing an object on the table triggers recognition of the object and its pose and displays arrows for where to measure, so measurement proceeds without consulting manuals or supervisors (\humanAugmentation).
Recognition does not deskill: no worker could learn to tell thousands of near-identical parts apart, so the model takes over a lookup rather than a craft, and workers gained independence in planning their measurement tasks (\emancipatoryDemocracy).
Reaching that performance required \specialization{} through synthetic training data generated by varying pose and background on CAD renderings of each critical object, and the prototype combined tangible and visual interaction (\multimodality).

Participation shaped problem definition and both the physical and on-screen interface; the AI methods were chosen by the researchers alone.
User rejection served as a design signal, but the democratic scope stayed shallow, worker access was scarce, and no longitudinal evaluation followed.

Industrial environments today face dual pressures:
Frequent job transitions requiring rapid skill acquisition and the need to reduce tedious work that drives employee turnover.
\textit{Quality assurance} (QA) exemplifies these challenges, demanding that workers quickly detect production errors and make precise measurements under time pressure.
This case study examines how LEGO Group QA workers participated in designing AI-enhanced Augmented Reality systems to support their measurement tasks, revealing how participatory approaches can integrate advanced computer vision with existing workplace practices rather than imposing external technological solutions.

\begin{figure}[htb]
    \centering
    \subfloat[]{\includegraphics[height=2.8cm]{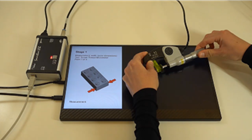}}
    \subfloat[]{\includegraphics[height=2.8cm]{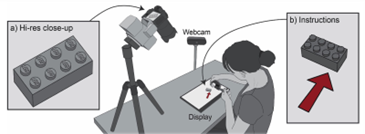}}
    \caption{\textbf{Industry object recognition.}
    Overview sketch of first version.
    Recognition of objects and their pose is detected by deep learning.
    The table surface display was selected to support work ergonomics.}
    \label{fig:manufacturing}
    \Description{An image and an illustration from the case study side by side. The image on the left shows a participant is following instructions from a display integrated into the table on which length of the test object to measure. The illustration on the right shows the study setup: A participant is seated at a table which has a builtin display showing instructions. Multiple cameras including a hi-res close-up camera capture the object pose, which is used to make instructions tailored to the object's current pose.}
\end{figure}

\subsubsection{Problem Domain}

Industry has a high frequency of shifting jobs, thus there is a need to support learning to perform new tasks.
There is also a concern to reduce more tedious aspects at the jobs to avoid such job shifts.
The specific challenge of industrial quality assurance (QA) is to support workers in making measurements and detecting production errors quickly to make necessary adjustments to the production process~\citep{BeloFenderFeuchtner2019, BeloWissingFeuchtner2023}.
To guide the QA workers in this case, we report on a case study where a Deep Learning (CNN-based) AI model was developed to process real-time video streams from high-resolution cameras to support visual recognition of industrial objects and their pose.
The studies lead to several cooperative prototyping experiments conducted between designers and users.

\begin{figure}
    \centering
    \includegraphics[width=0.8\linewidth]{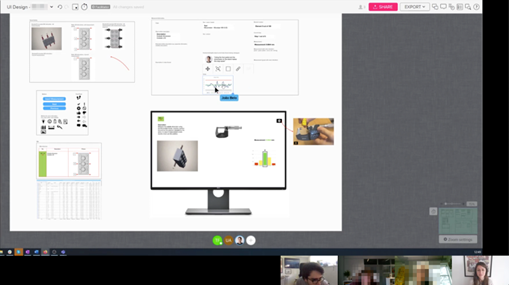}
    \caption{\textbf{Remote cooperative prototyping.}
    The tool was developed as part of the project to enable remote experimentation on the visual interface during the enforced social distancing of the COVID-19 pandemic.}
    \label{fig:remote-prototyping}
    \Description{A screenshot from an online design meeting. The shared view displays notes and ideas for the design of the case study.}
\end{figure}

In this case study, QA workers from LEGO Group were involved in Participatory Design activities.
The team conducted field studies at their workplace, as well as workshops on their needs for support.
Workplace visits, interviews as well as brainstorming inspired by future workshops~\citep{Boedker1995, FutureWorkshops} were applied to identify problems and envision solutions.
The ideas of using AI-based AR to guide measurements came out of these early activities. 
Experiments were conducted on both physical setups, interaction techniques, and visual feedback (Figure~\ref{fig:manufacturing}).
The project was later disrupted by COVID-19 lockdowns; hence a tool (Figure~\ref{fig:remote-prototyping}) was developed for conducting PD sessions remotely. 

\subsubsection{PAI Principles}

During PD workshops with the QA workers, the designers learned about the QA processes and the different artifacts involved in the procedures (\mutualLearning 
).
The ideas of using AI-based AR to guide measurements came out of these early studies.
The designers then demonstrated different Augmented Reality interfaces (glasses, projections, displays) with potentials for guiding the QA process. 
The workshops revealed that a main source of errors was coupled to QA measuring of near-symmetrical objects.
In addition, the QA workers rejected the AR glasses and pointed to traditional computer displays.
Experiments eventually lead to the choice of a display integrated into the table surface.
As part of the process it was important to consider both the current and future \artifactEcologies since the solution was dependent on both the CAD system that would deliver drawings of bricks and the database containing the actual quality assessment.
Hence possible solutions would be need to be developed in concert with these.

To explore \futureAlternatives, the second version of the prototype was made much simpler in the setup with only one high-res camera in an iPhone/iPad on a stationary pod to track objects on the surface~\citep{BeloWissingFeuchtner2023}.
AR glasses could still be an alternative in some future, but it turned out to be much more challenging for CNN algorithms to recognize objects and pose when objects are held in hand instead of sitting steady on the table surface.

QA workers were found to experience less need to consult manuals and supervisors on resolving issues regarding where to measure (\emancipatoryDemocracy).
Making the system guide the points of measurements does not take away skills from the users; as they would never learn to distinguish between the thousands of objects that look very similar.
Hence the AI recognition relieved them from constant manual look ups.
The QA support utilizes CNN models to resolve detailed distinctions in objects to be inspected.
Thus the QA workers could prioritize and plan their measurement tasks much more independently and without time consuming consulting of paper-based descriptions.
This would reduce the routine feeling of the tasks.
While there are no deep democracy aspects in this case, both workers and the company would benefit from fewer errors and more flexible work conditions.

\subsubsection{Design Challenges}

Since near-symmetrical objects were a challenge, the development focused on being able to detect the proper pose of such objects for measurements (\specialization).
To improve CNN performance on these objects, a lot of synthetic training data was generated by changing the pose and backgrounds of a few pictures (generated from CAD) of each of the critical objects.
In the end, the QA prototype provided both tangible and visual interaction using CNN-based object recognition (\multimodality).

Putting the object on the measurement table automatically triggered a recognition of an object and its pose, and in turn showed guiding arrows for where to measure. 
Thus the measurement process could potentially become a much more fluent and flexible task (\humanAugmentation), not interrupted by time-consuming look-ups in manuals and calls for supervisors to help decide on the pose of the object.

\subsubsection{Lessons Learned}

The PD activities focused on problem identification and the user interface design.
The problem identification formulated an AI challenge for the researchers to be able identify the right pose of near symmetrical and very similar objects to guide measurements.
The technical AI methods were chosen solely by the researchers, but both the physical and visual user interfaces were designed cooperatively.
Finally, it was a challenge to get access to industrial workers for creative processes due to shortage of workers and time constraints.

%% file: content/05-discussion.tex
\section{Discussion}
\label{sec:discussion}

The resource constraints that recur across the cases (Section~\ref{sec:synthesis}) align with Bødker and Kyng's emphasis~\citep{DBLP:journals/tochi/BodkerK18} on prototyping as essential for making participatory design work: early collaborative prototyping investments pay off through better alignment with actual practices.
Participants need support from their organizations and communities including, e.g., time to participate~\citep{bodker1991design}.
Organizational scale determines feasibility: LEGO Group could invest in extended worker collaboration while the startup needed more agile approaches.
The complex levels of participation and need for resources at each level is discussed further by \citet{TyingKnots}. 
The pattern suggests participatory AI needs methodological diversity: both deep approaches for well-resourced contexts and lighter methods for constrained environments, matching participatory intensity to organizational capacity.

One such lighter approach are participatory data practices under \humanAugmentation.
In this spirit, the agricultural example offers volunteering of training data to finetune the AI model towards the needs of the farmers sharing their data.
Similarly, the educational example mentions teachers uploading additional datasets and sharing educational practices around their use of AI tools, which is yet another lightweight form of participating in the discussion around and evolution of the used tools if in-person workshop participation is not economic or possible.

\subsection{Cultural Specificity and the Need for Methodological Pluralism}

While this paper advocates for participatory approaches to AI development, we must acknowledge that 
grounding our framework in a PD tradition that arose in 1970s Scandinavia risks perpetuating the very power imbalances we seek to address.
That tradition emerged from specific historical conditions (strong labor unions negotiating workplace automation~\citep{KyngMathiassen1982, EhnSandberg1983, Nygaard1978}, social democratic governance structures, and relatively homogeneous societies) that do not exist universally, or maybe not even at all in the cultural mosaic of our present world.
Canonizing these particular practices as exemplary democratic design may inadvertently impose a Western, white-dominated (WEIRD~\citep{DBLP:conf/chi/SturmOLADR15}) cultural framework onto diverse global contexts, contradicting core principles of postcolonial HCI~\citep{DBLP:conf/chi/IraniVDPG10} that challenge such universalizing tendencies.
The irony is stark: advocating for democratic participation while prescribing a culturally specific approach developed in one of the world's wealthiest regions fundamentally undermines the inclusive values we put forth.

Critical PD scholarship has documented these exclusivity problems in detail.
\citet{DBLP:journals/pacmhci/HarringtonEP19} argue that PD as instantiated in the design workshop is ``an affluent and privileged activity'' that neglects challenges faced by underserved populations, including historical context, community access, and the potential for unintentional harm when collecting personal narratives.
\citet{DBLP:conf/chi/PierreCCPP21} identify how data-centered participatory research can place epistemic burdens on minoritized groups---forcing communities to justify their experiences in terms legible to researchers---even in projects focused on social justice outcomes.
\citet{DBLP:journals/pacmhci/ElsayedAliBC23} find that even practitioners focused on inclusion face persistent challenges: creating shared spaces that genuinely empower partners, developing common ground among stakeholders with different expertise, and balancing funding requirements with the open-ended nature of authentic PD.
These critiques apply with greater force to AI, where technical complexity compounds existing barriers to participation.

This tension suggests framing our contribution properly.
Rather than promoting PAI as a universal solution, we present it as one culturally situated response to technological automation that offers insights for broader democratic engagement with AI systems.
In the spirit of PD, we instead offer alternatives that help question what is often mainstream AI practices.
However, meaningful participation in decisions affecting one's work and life, transparency in algorithmic systems, and preservation of human agency transcend any particular cultural implementation.

Truly democratic AI development requires drawing from a whole host of diverse perspectives.
No single methodology, whether Scandinavian PD or otherwise, can address the complex sociotechnical challenges of AI development.
Instead, HCI design should be indigenous~\citep{Duarte2017, SmithWinschiers2020, Kumar2021} in the deepest sense: emerging from and accountable to the specific communities and contexts where technologies will be deployed.
\citet{DBLP:conf/fat/HutchinsonLCC25} demonstrate what this can look like in practice through their work on speech technologies for Australian Aboriginal English, where culturally appropriate and participatory processes were integrated throughout a project that recognized both opportunities and risks of AI development for Indigenous communities.

\subsection{Values, Power, and Democratic Participation}

Current AI development embeds values through opaque corporate processes that masquerade as technical neutrality.
Who decides what counts as valid knowledge?
Which communities benefit from AI models and which bear the costs?
How are value trade-offs resolved when AI systems must choose between competing priorities?
These are inherently political decisions currently delegated to corporate boardrooms and technical teams without democratic input.

Several existing topics within HCI can help here.
Values in design research~\citep{Friedman2013} demonstrates that every technology encodes specific priorities, yet AI systems obscure these value choices behind claims of algorithmic objectivity.
Critical computing~\citep{Agre1997, Winner1980} exposes how seemingly neutral systems perpetuate existing power structures and inequalities.
The challenge for participatory approaches lies in developing processes for democratic value articulation that can compete with the speed and scale of corporate AI development.

Feminist HCI~\citep{Bardzell2010} and care-oriented design~\citep{Puig2020} offer additional nuances. 
Feminist computing traditions emphasize collaborative knowledge creation, attention to marginalized voices, and recognition that technology shapes social relations in profound ways.
Current AI systems often embody masculine-coded assumptions about efficiency, optimization, and individual agency that ignore interdependence and collective flourishing.
How can participatory approaches integrate care ethics and relational perspectives that challenge productivity-focused framings of AI?
Can our democratic processes adequately represent care work that AI models increasingly affect?

\citet{DBLP:conf/fat/SureshMDBCCTSD22} demonstrate that feminist participatory approaches can be operationalized from the outset rather than retrofitted.
Their work on feminicide counterdata collection applied intersectional feminist goals throughout the ML pipeline: from problem conceptualization to data collection to model evaluation.
Their methodological contributions---iterative annotation processes that interrogate framing concepts, models that focus on intersectional identities rather than statistical majorities, and multi-step evaluation combining quantitative and qualitative elements---show how participation can be structured to challenge rather than reinforce existing power relations.

Additionally, entanglement HCI~\citep{Frauenberger2019} and more-than-human perspectives extend human-centered design to consider multi-species and non-humans such as AI agents~\citep{Yoo2023, Nicenboim, Eriksson2024}.
This challenges the notion of products being ``industrial artifacts'' to ``fluid assemblages,'' which are networked, dynamic, and with constantly evolving forms and functions depending upon context~\citep{redstrom2018changing}.
So how can PD take account for such multi-faceted, fluid, and rapidly developing forms and functions and the challenges that our intimate entanglement with non-humans imply?

Yet participation alone may not be sufficient when power imbalances are severe.
\citet{Agnew2023Resistance} develop a theory of ``technologies of resistance'' that complements participatory frameworks.
They argue that affected communities sometimes need tools to resist AI systems: to re-align values to local contexts, shift power from model owners to data subjects, and enable agency that participation cannot provide when designers hold structural advantages.
This is not a rejection of participation but a recognition of its limits: resistance becomes necessary when participatory mechanisms are absent, co-opted, or insufficient to address entrenched inequalities.
Relatedly, \citet{DBLP:conf/chi/AgnewBCDEPMM24} warn against the ``illusion of artificial inclusion'': proposals to replace human participants with AI surrogates that promise diversity and reduced costs but conflict fundamentally with the values of representation, inclusion, and understanding that justify participation in the first place.
PAI must remain vigilant against both the co-optation of participation and its simulation.

\subsection{Civic and Social AI}

Participatory Design emerged from workplace struggles, yet AI's impact extends far beyond traditional employment contexts.\footnote{History is repeating itself even here as PD also quickly started to envelop everyday and recreational activities~\citep{DBLP:conf/nordichi/Bodker06}.}
Students encounter AI in educational platforms that can perpetuate bias against non-native English speakers and underrepresented groups~\citep{Baker2022}, patients rely on AI-assisted diagnosis that may underdiagnose conditions in marginalized populations~\citep{Obermeyer2019, Seyyed-Kalantari2021}, citizens interact with AI in government services through algorithmic governance systems~\citep{Margetts2015}, and communities navigate AI-mediated social platforms.
Each domain presents distinct challenges for democratic participation that require different approaches to civic engagement~\citep{Fung2006}.

How do we scale democratic engagement beyond workplace contexts for which PD was originally intended?
What forms of participation are appropriate for different domains: individual choice, community governance, or democratic representation?
The intensity and commitment that characterized Scandinavian workplace democracy may not translate to civic contexts where stakes and relationships differ fundamentally.

\citet{DBLP:journals/firstmonday/YoungESTGHM24} argue that these tensions between participation and scale are contingent rather than irresolvable.
They observe that scale refers not just to system size but to the infrastructural investments needed to extend a system across contexts.
Just as scaling commercial AI required significant investment, scaling participation will require dedicated infrastructure for shifting power---what they call the ``practical dimension'' of the participatory tradition's commitments.
This reframes the challenge: rather than asking whether participation can scale, we should ask what infrastructure would enable it to do so.
\citet{DBLP:conf/emnlp/NekotoMMFFAMKOS20} provide an existence proof by demonstrating that participatory research for machine translation can produce datasets and benchmarks for over 30 African languages by enabling contributors without formal training to make scientific contributions.


\subsection{Conditions for Participatory AI}
\label{sec:conditions}

So far in this paper, we have described what participatory AI should look like.
A harder question is how to bring it about.
It is clear from current developments in industry and society that PAI will not emerge spontaneously from the competitive dynamics of commercial AI development, where speed to market and model scale dominate incentive structures.
Nor is it realistic to expect that voluntary adoption by technology companies will produce systemic change.
If participatory AI is to move beyond isolated research projects, it requires structural conditions---legal, institutional, and organizational---that create space for democratic engagement with AI systems.

\paragraph{Regulatory frameworks.}

The EU's AI Act~\citep{AIAct2024} and the General Data Protection Regulation (GDPR)~\citep{gdpr} represent advanced regulatory infrastructure for AI governance, yet neither was designed with participation in mind.
The AI Act operates in the tradition of product safety law: its requirements are largely \textit{ex ante}, governing what happens before an AI system reaches the market, and its regulated actors are providers, deployers, importers, and distributors along the value chain~\citep{AIAct2024}.
End-users and affected communities are not included as core actors with agency in this framework.
For example, their participation is limited to the right to lodge a complaint with a regulator if they think the law is breached (Art.~85) and the right to a ``clear and meaningful explanation'' when directly affected by AI-assisted decision-making (Art.~86).
A third provision---requiring organizations to ensure ``a sufficient level of AI literacy'' among staff operating AI systems (Art.~4)---touches on \mutualLearning, though without any requirement for mutuality on the part of the person affected by the system output.
The provisions on testing AI systems in real-world conditions (Art.~60--62) require informed consent from test subjects, but frame participation as a safeguard rather than a design resource.

The GDPR, as a fundamental rights law, does place people at the center rather than the value chain, yet it has often been criticized for its individualistic orientation~\citep{viljoen2021relational, MahieuAusloos2020}.
Its rights---to information, access, rectification, erasure, objection, and human intervention---extend only to one's own personal data and do not prescribe collective participatory mechanisms.
There are hints of a collective imaginary, however.
Art.~80 allows an individual ``to mandate a not-for-profit body, organisation or association'' to lodge complaints and demand judicial remedies on their behalf.
Although it is a stretch to call this participation, this article is increasingly being used in collective actions for compensation.
There is also an underused provision that deserves attention: when processing involves new technologies or is otherwise high-risk, controllers must conduct a Data Protection Impact Assessment (DPIA, Art.~35), and the GDPR states that ``where appropriate, the controller shall seek the views of data subjects or their representatives on the intended processing.''
The clause ``where appropriate'' gives considerable interpretative leeway to the controller, but it constitutes a latent hook for participatory processes that could, with strategic interpretation, be expanded to support community involvement in AI design decisions.

In sum, existing EU regulation contains embryonic participatory elements but falls short of creating conditions for the kind of democratic engagement our framework envisions.
Furthermore, recent developments suggest these conditions may be deteriorating: the EU has already begun rolling back provisions of the AI Act through the Omnibus simplification package, driven by concerns about competitiveness relative to US and Chinese AI development~\citep{EUOmnibus2025}.

\paragraph{Collective bargaining and labor institutions.}

The original Scandinavian PD emerged because strong trade unions negotiated over the terms of workplace automation~\citep{KyngMathiassen1982, EhnSandberg1983, Nygaard1978}.
Today, this institutional infrastructure has weakened considerably.
While the EU's Platform Work Directive grants platform workers rights relative to the algorithmic systems they work under, system design itself is largely absent from collective bargaining agreements.
This represents a missed opportunity: if trade unions and professional associations made AI system design a subject of negotiation---as their Scandinavian predecessors did with computer systems in the 1970s---it would create an institutional basis for participatory AI that does not depend on the goodwill of technology providers.
Perhaps it is time for labor unions to go once more unto the breach in defense of average citizens.

\paragraph{Implications.}

These observations point to a gap between the participatory ideals we advocate and the institutional infrastructure available to support them.
Regulatory frameworks treat affected communities as passive recipients of safe products rather than active participants in design.
Labor institutions have not yet adapted to make AI system design a subject of democratic negotiation.
Addressing this gap does not require entirely new legal instruments; rather, it requires expanding existing mechanisms---impact assessments, literacy requirements, collective bargaining---to include genuine participatory elements.
Our case studies illustrate that such participation produces better outcomes when it does occur.
The challenge is creating the structural conditions that make it systematic rather than exceptional.

%% file: content/06-research-vision.tex
\section{Future Research Directions}
\label{sec:research-vision}

Our framework and the cross-case synthesis (Section~\ref{sec:synthesis}) point to changes needed across how communities engage with algorithmic systems.
Yet realizing this vision requires systematic changes across multiple dimensions: from the technical architecture of AI systems to the governance structures that guide their development and deployment~\citep{DBLP:journals/tochi/BodkerK18, DelgadoYangMadaio2023}. 
This section outlines a research agenda that bridges immediate practical needs with longer-term, transformative goals.

\newcommand{\grounding}[1]{\hfill{\footnotesize\textit{Grounding: #1}}}

\subsection{Design Recommendations}
\label{sec:d-recommendations}

Current AI development prioritizes technical performance metrics---accuracy, efficiency, scale---over alignment with human practices and values~\citep{Bommasani2021, rudin2019stop}.
This design philosophy produces systems that may excel at benchmark tasks while failing to support the nuanced, contextual work that defines human expertise~\citep{DBLP:conf/chi/DoveHFZ17}.
Participatory AI demands different design priorities: systems that learn from and with communities, interfaces that preserve rather than automate away human judgment, and architectures that can adapt to local needs without sacrificing broader interoperability.

\begin{enumerate}[label=DR\arabic*]

    \item\textbf{From Universal Tool to Situated Collaborator.}
    A central tenet in participatory AI is that people are experts in their own practices.
    This situated expertise and agency is often lost in current universalist framings of AI systems and tools.
    When models automate parts of writing, reporting, programming, or knowledge synthesis, the encoded expertise often reflects generalized datasets and distant value judgments rather than local ways of knowing and doing.
    We therefore argue for AI systems as \textit{situated collaborators}: systems co-designed and fine-tuned with practitioners who understand domain nuances, edge cases, and the social meanings of ``good'' performance.
    This reframing shifts participation to earlier stages: rather than tweaking interfaces, stakeholders co-author the system's goals, guardrails, evaluation criteria, and integration with local knowledge bases, with local appropriateness as a primary design goal.
    \grounding{P1, P3; education case}
    
    \item\textbf{Cooperative Human–AI Interaction.} 
    Positioning AI systems as collaborators foregrounds cooperative interaction as a site of mutual learning. 
    The design of AI systems should encourage and facilitate learning about AI in relation to the specific domain by the potential users so that the design process not only leads to creating a system, but also provides a learning opportunity for both designers and users.
    Modern interfaces for interacting with particularly generative AI models hide the complexity of how an answer to a prompt is produced as coming from a single ``intelligence,'' where in reality it is a complex interplay between a model and myriads of tools and auxiliary data sources. 
    AI interfaces should make this ensemble legible: which model, which tools, which retrieved sources shaped a given output.
    \grounding{P1; creativity case}
    
    \item\textbf{Archipelagos of Local AI Sovereignty.} 
    Rather than monolithic global systems, we advocate for networks of locally-controlled AI models that preserve cultural diversity and community self-determination while enabling beneficial knowledge sharing.
    At the same time, there are situations where access to broader knowledge is necessary.
    This may involve local models getting users' consent to query larger models as in the hybrid approach in Apple Intelligence, or to query other local models within the same federated ecosystem, for example a municipal school's model querying that of the ministry of education.
    \grounding{P3}
    
\end{enumerate}
    
\subsection{Sociotechnical Recommendations}
\label{sec:st-recommendations}

Technology is never politically neutral~\citep{WinogradFlores1986}.
AI systems encode particular visions of how work should be organized, how decisions should be made, and who deserves to participate in shaping our collective future~\citep{DBLP:conf/eaamo/BirhaneIPDEGM22}.
The current trajectory concentrates power while distributing risk, driven not by corporate actors alone but by an alignment of commercial and political interests that pursues deregulation where it enables extraction and state intervention where it protects capital, whether through public subsidies for private AI infrastructure or geopolitical maneuvering to secure material resources.
Participatory AI requires governance structures that reverse this dynamic: democratic oversight of systems that shape public life, community ownership of AI tools, and institutional mechanisms that ensure those affected by algorithmic decisions have control over them~\citep{SloaneMossAwomolo2022}.

\begin{enumerate}[label=STR\arabic*]


    \item\textbf{Reconfiguring Practice, Power, and Ownership.}
    Introducing AI systems reshapes cooperative routines and power relations both within and beyond workplaces: who controls task definitions and work practices, how cognitive load is redistributed, and where authorship and accountability now sit.
    PD holds that those affected by technology should control that technology.
    This means workers owning the AI tools that augment their labor, communities governing the AI systems that shape their civic life, and students controlling educational AI rather than becoming subjects of surveillance through their data.
    Ownership implies not only decision rights but a claim on the value derived from community data and expertise, which AI development currently extracts to benefit distant shareholders.
    Two existing levers can carry this (Section~\ref{sec:conditions}): making AI system design a subject of collective bargaining, as the trade unions of the 1970s did with computer systems, and using the GDPR's impact-assessment consultation clause (Art.~35) to require that affected communities are heard before high-risk processing begins.
    \grounding{P4; agriculture case; Section~\ref{sec:conditions}}

    \item\textbf{Defending Future Alternatives.}
    PD's commitment to future alternatives treats development as an open and dynamic exploration of the roles AI might play in supporting human practices.
    Here, empowerment is central to the community's enhanced capacity to pursue goals, and to revise or refuse technologies that conflict with these.
    New technological potentials must be explored systematically and with curiosity, but the tech industry's mantra of \textit{``Move Fast and Break Things''} must be resisted, particularly in areas where lives are at stake when AI systems increasingly affect healthcare, education, and justice.
    \grounding{P2; creativity and knowledge cases}
    
    \item\textbf{AI as Public Infrastructure, Not Private Product.}
    No proprietary algorithm should shape public life without public oversight.
    We envision AI systems governed like public utilities: transparent, regulated, and democratically accountable.
    Laws such as the EU GDPR~\citep{gdpr} and AI Act~\citep{AIAct2024} show that political bodies can impose transparency requirements, even though neither yet creates conditions for participation (Section~\ref{sec:conditions}).
    Local initiatives, such as the Swiss Apertus\footnote{\url{https://ethz.ch/en/news-and-events/eth-news/news/2025/09/press-release-apertus-a-fully-open-transparent-multilingual-language-model.html}} and the Dutch GPT-NL\footnote{\url{https://gpt-nl.nl/}} models, demonstrate that researchers can support these principles by developing open and transparent models that respect legal safeguards, include underrepresented cultures, and make training data and weights publicly available. 
    However, we also recognize potential tension in making training data openly available.
    As our agricultural use case illustrates (see Section~\ref{sec:agriculture}), data sharing may not always be possible or desirable.
    Transferring data ownership to private companies may be necessary to generate this data in the first place, thus offsetting hardware costs and enabling companies to create such AI models.
    \grounding{P4; knowledge case}

\end{enumerate}

\subsection{Open Research Questions}

While our case studies and recommendations provide concrete pathways forward, they also reveal fundamental tensions that require sustained research attention.
The questions arise from the inherent complexity of designing systems that must simultaneously serve local communities and operate at global scale, preserve human expertise while leveraging computational power, and maintain democratic accountability in rapidly evolving technological landscapes~\citep{SureshTsengYoung2024, DBLP:conf/eaamo/BirhaneIPDEGM22}.
These research directions demand interdisciplinary collaboration between computer scientists, social scientists, designers, lawmakers, and the communities most affected by AI systems~\citep{DBLP:series/synthesis/2022Bodker}.
Progress will require not just technical innovation but new institutional forms, governance mechanisms, and ways of organizing collaborative research that embody participatory principles in practice.

\begin{enumerate}[label=ORQ\arabic*]

    \item\textbf{Scale versus Situatedness.}
    Models operate across cultures at very large scale, while PD centers local specificity.
    \citet{DBLP:journals/firstmonday/YoungESTGHM24} reframe this as an infrastructure question (Section~\ref{sec:discussion}); what remains open is which coordination mechanisms preserve diversity without sacrificing interoperability, and who builds and pays for them.

    \item\textbf{Expertise Distribution.}
    The technical complexity of AI systems can widen epistemic gaps.
    Mutual-learning approaches through training in AI literacy for stakeholders, and domain literacy for developers are needed to ensure democracy is maintained. 

    \item\textbf{Temporal Dynamics.}
    Unlike static software or mechanical automation, AI systems can change rapidly.
    Participatory processes must therefore be continuous. 
    We should not only encourage but demand the oversight of significant changes to AI systems, especially in critical domains.
    This will require developing guidelines and/or checklists to facilitate the examination of the changes.
    Mutual learning and empowerment of domain experts will ease this process by giving them the tools to oversee and examine the changes themselves.
    
\end{enumerate}

The path toward participatory AI is neither inevitable nor impossible.
The history of PD shows that workers and communities can successfully negotiate with powerful technologies when equipped with democratic tools and institutional support.
Today's AI moment presents a similar opportunity: to ensure that the systems reshaping our world amplify human agency rather than diminish it.
By embedding participatory principles into AI from the ground up, we can create technologies that serve communities rather than extract from them, that enhance human expertise rather than replace it, and that strengthen democracy rather than undermine it.
The technical foundations exist.
The democratic precedents are proven.
What remains to be seen is whether we possess the collective will to choose participation over extraction, empowerment over automation, and human flourishing over algorithmic efficiency.

%% file: content/10-acks.tex
This work was supported partly by Villum Investigator grant VL-54492 by Villum Fonden.
Any opinions, findings, and conclusions expressed in this material are those of the authors and do not necessarily reflect the views of the respective funding agencies.